\documentclass{article}

\usepackage{arxiv}

\usepackage{amsmath}
\usepackage{xcolor}
\usepackage{algorithmic}
\usepackage{array}
\usepackage{todonotes}
\usepackage{xspace}
\usepackage{booktabs}
\usepackage{graphicx}
\usepackage{subcaption}
\usepackage{caption}
\usepackage{tabularx}

\usepackage{hyperref}
\usepackage[table,dvipsnames]{xcolor}

\colorlet{mylinkcolor}{YellowOrange}
\colorlet{mycitecolor}{Aquamarine}
\colorlet{myurlcolor}{violet}
\hypersetup{ 
        colorlinks=true, 
        linkcolor=mylinkcolor, 
        citecolor=mycitecolor, 
        urlcolor=myurlcolor, 
        filecolor=myurlcolor }

\usepackage{cleveref}
\usepackage{url}

\newcommand{\name}{\textsc{DeepChronos}\xspace}
\newcommand{\ourparagraph}[1]{\noindent\textbf{#1}}

\usepackage[most]{tcolorbox}
\newtcolorbox{takeaways}{
    colback=gray!5,
    colframe=black,
    title=Takeaways
}

\definecolor{bordeaux}{RGB}{128, 0, 32}

\usepackage[inline]{enumitem}
\newenvironment{inlinelist}{\begin{enumerate*}[label=\emph{(\roman{*})}]}{\end{enumerate*}}

\usepackage{fontawesome}
\newcounter{takeawaycounter}

\title{\name{}: Temporal Neuro-Symbolic Reasoning for Multi-Step Attack Detection}

\author{
  Sofia Burkow\textsuperscript{1} \quad
  Andrea Agiollo\textsuperscript{1} \quad
  Enkeleda Bardhi\textsuperscript{1} \quad
  Fernando Kuipers\textsuperscript{1} \\
  \bfseries
  Kaitai Liang\textsuperscript{2} \quad
  Georgios Smaragdakis\textsuperscript{1} \\[1.5ex]
  \textsuperscript{1}Delft University of Technology \qquad
  \textsuperscript{2}University of Turku \\[1ex]
  \texttt{sofiaburkow@proton.me}, \texttt{kaitai.liang@utu.fi}, \\ \texttt{\{A.Agiollo-1, E.Bardhi-1, F.A.Kuipers, G.Smaragdakis\}@tudelft.nl}
}

\date{}

\renewcommand{\shorttitle}{\name{}}

\hypersetup{
pdftitle={A template for the arxiv style},
pdfsubject={q-bio.NC, q-bio.QM},
pdfauthor={David S.~Hippocampus, Elias D.~Striatum},
pdfkeywords={First keyword, Second keyword, More},
}

\begin{document}
\maketitle

\begin{abstract}
Modern cyber threats commonly unfold as multi-step attacks comprising temporally and causally related phases. 
Current defenses commonly follow a \emph{detect first and correlate later} paradigm.
Intrusion detection systems classify events independently, while downstream systems reconstruct attack progression from the resulting alerts or audit records.
In this context, learning-based detectors learn attack-step dependencies implicitly as statistical regularities in the data and cannot guarantee that their predictions follow the real attack progression.
Conversely, correlation systems represent attack relationships explicitly, but apply this knowledge only after the underlying attack observations have been detected. 
We present \name{}, a neuro-symbolic network intrusion detection system that shifts the paradigm from \emph{detect first and correlate later} to \emph{detection through logical progression}. 
\name{} combines learned flow-level evidence with explicit attack-phase rules, temporal prerequisites, and previously inferred phases, allowing the evolving attack history to inform and constrain each detection decision.
Evaluations on DARPA 2000 and AIT-LDS V2 show that \name{} improves macro F1 over purely neural baselines by up to $11.3\%$, eliminates causal violations under strict temporal constraints, and maintains millisecond-scale inference latency. 
These results demonstrate that integrating explicit attack progression knowledge into detection produces more accurate and temporally coherent multi-step attack predictions.
\end{abstract}


\section{Introduction}\label{sec:intro}

Modern cyber-attacks rarely consist of a single malicious event.
Adversaries pursue their objectives through interdependent steps, such as reconnaissance, initial access, privilege escalation, lateral movement, and more as reflected in the Cyber Kill Chain and MITRE ATT\&CK~\cite{hutchins2011intelligence,MITREATTACK}.
Detecting these multi-step attacks (MSAs) is challenging because their actions may be distributed across hosts, separated in time, and insufficiently suspicious when considered in isolation~\cite{ning2002constructing,haas2019alertCorrelation}.
Together, however, they can cause severe harm.
For example, in the SolarWinds incident, the adversary combined supply-chain access with credential and identity abuse, persistence, lateral movement, collection, and exfiltration~\cite{mitreSolarWinds}.
More recently, in the 2024 Change Healthcare intrusion, attackers used compromised credentials to access a remote portal without multifactor authentication, moved laterally, exfiltrated data, and deployed ransomware nine days later~\cite{witty2024changeHealthcare}.

Both incidents show how the final impact was preceded by a progression of consequential activities, illustrating why detecting only the terminal event is insufficient for effective defense.
Moreover, detecting an MSA early requires recognizing not only its constituent steps, but also the temporal and causal relations connecting them.
For example, lateral movement generally presupposes an established foothold or valid credentials, while exfiltration presupposes access to relevant data.
Although attackers may skip, repeat, or interleave activities, attack steps remain subject to prerequisite and partial-order relationships~\cite{hutchins2011intelligence,ning2002constructing}.
These relations allow weak observations to reveal a developing intrusion.
Conversely, predictions such as exfiltration before compromise may be locally plausible yet inconsistent with the evolving attack context.
An effective detector should therefore use previously observed steps to guide each new prediction and identify the attack before its objective is reached.

Most existing defenses against MSAs follow a \emph{detect first, correlate later} paradigm.
In the first stage, an IDS examines network or system observations and produces independent alerts.
Signature- and rule-based IDSs detect known behaviors using expert-defined patterns, thus typically recognize atomic actions rather than attack progressions~\cite{scarfone2007guide,sommer2010outside}.
Meanwhile, machine learning (ML)-based IDSs provide greater flexibility by learning malicious or anomalous behavior from data~\cite{mirsky2018kitsune,vinayakumar2019deep,agiollo2023gnn4ifa,bardhi2025mine} and aiming to learn attack-step dependencies implicitly as statistical regularities---e.g., using recurrent architectures such as LSTMs~\cite{hochreiter1997long} and GRUs~\cite{cho2014learning}.
However, because these dependencies are only learned as statistical correlations, the model may predict phases whose prerequisites have not been observed and produce invalid attack progression predictions.
%
%
%

In the second stage, alert-correlation systems attempt to recover the attack progression by grouping alerts and reconstructing higher-level attack scenarios~\cite{ning2002constructing,haas2019alertCorrelation,wang2021maac}.
These systems make attack-step relationships explicit and can support both online analysis and post-incident investigation.
Nevertheless, their reasoning remains downstream of detection.
The alerts have already been generated before the attack context is considered.
Consequently, attack progression knowledge cannot influence how each event is classified and false alerts may produce spurious attack paths.
The conventional pipeline therefore exposes a fundamental separation: detection systems process individual events without explicit attack steps knowledge, while correlation systems reason about attack steps only after the detection decisions have been made.

We propose to overcome this separation by explicitly enforcing the logical attack progression knowledge within the detection system---i.e., \emph{detection through logical progression}.
Rather than learning attack-step relationships only implicitly or applying them only during downstream correlation, progression knowledge should directly inform the detection of each new observation.
To this end, we present \name{}, the first neuro-symbolic IDS that embeds explicit MSA progression logic directly into attack detection.
Neuro-symbolic systems combine the representation-learning capabilities of neural networks with explicit reasoning of symbolic AI~\cite{garcez2019neurosymbolic,marra2024survey,ciatto2024symbolic,saad2026deepweave}.
Similarly, \name{} uses a neural component to identify malicious behavior from noisy network-flow data and a symbolic reasoning engine to encode attack-phase rules and temporal prerequisites of MSAs.
For each flow, it reasons jointly over learned evidence and the previously observed history of the relevant attack context, allowing earlier steps to inform and constrain subsequent predictions.
A lightweight memory maintains separate histories for active contexts, supporting attacks that unfold over extended periods.
Unlike conventional IDSs and post-hoc correlation systems, \name{} integrates uncertain neural evidence and explicit MSA logic within probabilistic attack-phase inference.
Thus, \name{} raises an alarm when the accumulated context satisfies a configurable multi-step attack criterion, enabling weak early evidence to contribute to detection while limiting alarms from isolated predictions.

This paper makes the following contributions:
\begin{itemize}
    \item We formulate MSA detection as context-dependent neuro-symbolic inference governed by explicit temporal and attack-phase knowledge.
    \item We design \name{}, which integrates neural flow classification, probabilistic logical reasoning, and context-specific attack memory in an end-to-end pipeline.
    \item We introduce metrics for causal and regression violations and evaluate \name{} on DARPA 2000~\cite{lippmann2000darpa} and AIT-LDS V2~\cite{landauer2021have,landauer2023maintainable} against neural, symbolic, and anomaly-detection baselines.
    \item We analyze \name{} data efficiency, cross-domain generalization, robustness to missed steps, and runtime overhead.
\end{itemize}

The remainder of this paper is organized as follows.
\Cref{sec:background,sec:threat_model} present the background and threat model, respectively.
\Cref{sec:design_details} describes \name{} architecture, while \Cref{sec:implementation} details its implementation and \Cref{sec:evaluation} elaborates on its evaluation.
Lastly, \Cref{sec:discussions_limitations} discuss limitations and \Cref{sec:related_work} the related work, while \Cref{sec:conclusions} concludes the paper.

\section{Background}\label{sec:background}


\subsection{Multi-Step Attacks}
A multi-step attack (MSA) comprises related actions performed toward a common adversarial objective~\cite{navarro2018survey}.
Attack frameworks describe these actions as phases such as reconnaissance, initial access, persistence, lateral movement, collection, and exfiltration~\cite{MITREATTACK,hutchins2011intelligence}.
We denote the modeled phases by $\mathcal{P}=\{p_1,\ldots,p_M\}$ and an attack progression by a sequence over $\mathcal{P}$.
A prerequisite relation $p_i \prec p_j$ indicates that $p_i$ must be observed before $p_j$ to be considered plausible.
Depending on the attacker, phases may form a strict sequence or there may exist omissions and phase repetitions, as real attackers may skip, repeat, or interleave activities.
MSA detection therefore requires both classifying individual phases and determining whether their accumulation indicates an ongoing attack.
Predicting a phase before a required predecessor constitutes a \emph{causal violation}, whereas predicting an incompatible earlier phase constitutes a \emph{regression violation}.

\subsection{Network Intrusion Detection Systems}

A network intrusion detection system (NIDS) monitors traffic for malicious activity or policy violations~\cite{scarfone2007guide}.
Signature-based NIDSs match expert-defined attack descriptions, while data-driven NIDSs learn malicious or anomalous traffic patterns~\cite{sommer2010outside}.
NIDSs may analyze packets or aggregate them into flows, typically identified by endpoint addresses, ports, and transport protocol.
Flow features summarize properties such as duration, traffic volume, directionality, and inter-arrival times, and remain available when payloads are encrypted.
A learned NIDS commonly estimates $p_\theta(Y=y\mid x)$ for a traffic observation $x$ and an attack class prediction $y$, while MSA detection additionally conditions phase inference on previous phases and attack-progression knowledge.

\subsection{Neuro-Symbolic Reasoning}
Neuro-symbolic AI combines neural representation learning with explicit knowledge representation and symbolic reasoning~\cite{garcez2019neurosymbolic,marra2024survey,ciatto2024symbolic,saad2026deepweave}.
Neural models extract uncertain patterns from data, while symbolic models express domain knowledge through facts and rules.

\paragraph{DeepProbLog.}
DeepProbLog~\cite{manhaeve2018deepproblog,manhaeve2021neural} extends the probabilistic logic language ProbLog with neural predicates~\cite{deRaedt2007problog}.
A logic program represents knowledge through \emph{facts}, \emph{predicates}, and \emph{rules}. 
A fact states that a relation holds, while a predicate represents a general relation. 
Rules derive new facts from existing knowledge.
For example, a rule may identify a scanning phase from the protocol and number of contacted ports. 
ProbLog assigns probabilities to uncertain facts, defining possible worlds in which different facts hold. 
A query's probability is then computed as the total probability of the worlds in which it can be proven.
DeepProbLog preserves this semantics while allowing selected probabilities to be produced by neural networks. 
A \emph{neural predicate} maps an input $x$, such as a network flow, to a probability distribution over symbolic values, such as benign and malicious. 
These values can participate in logical rules together with deterministic and probabilistic knowledge. 
To answer a query, DeepProbLog evaluates the required neural predicates and performs probabilistic inference over the logical proofs supporting the query.
Under end-to-end training, query-level gradients propagate through inference to the neural parameters, allowing learning and reasoning to be optimized jointly.

\section{Threat Model}
\label{sec:threat_model}

\subsection{Attacker Model}
\label{ssec:attacker_model}

We consider adversaries whose actions form a MSA progression consistent with established lifecycle models, such as the Cyber Kill Chain \cite{CYBERKILLCHAIN}. Although the concrete techniques used at each stage may vary, we assume that the attack retains a staged structure comprising reconnaissance, initial compromise, and post-compromise activities. Similar assumptions underlie systems such as HOLMES, which models coordinated activities including reconnaissance, command and control, privilege escalation, lateral movement, and exfiltration \cite{milajerdi2019holmes}.
Within the scope of this work, we further assume that the modeled attack phases produce observable network-flow characteristics. This assumption limits the system to the network-visible portion of an attack. The host activities that do not generate distinctive network traffic are outside the current detection scope. Finally, we allow related attack flows to be separated by extended intervals, ranging from hours to days. The system must therefore retain historical information beyond the neural input window to associate temporally distant observations. This is consistent with prior work showing that attack behavior may span multiple flows and extended time periods \cite{qu2026tracegram}, as well as with provenance-based systems designed to preserve historical context for long-running attacks \cite{han2020unicorn}.

\subsection{Evasion Assumptions} \label{ssec:evasion_assumptions} 
We consider non-adaptive adversaries whose actions preserve the network and temporal characteristics of the modeled multi-step attacks. Specifically, we assume that attackers do not possess detailed knowledge of \name{} or deliberately alter their behavior to evade its neural classification and symbolic correlation mechanisms. Potential adaptations include delaying, reordering, or omitting attack phases, etc. 
Such evasion strategies are feasible, as prior work has demonstrated practical problem-space attacks against learned security models and behavioral mimicry against provenance-based intrusion detectors \cite{pierazzi2020intriguing,goyal2023mimicry}. However, evaluating these strategies requires an adversarial methodology that preserves the attack objective and relevant network constraints. We therefore restrict the current evaluation to non-adaptive executions of the modeled attack scenarios and leave robustness against adaptive adversaries outside the scope of this work.

\subsection{Missed Attacks and False Alarms} \label{ssec:missed_attack_false_alarms} 
The objective of \name{} is to detect multi-step attacks before their final objective is reached. Early-stage activities, such as reconnaissance, are often ambiguous in isolation, but missing them may leave the attack history incomplete and hinder subsequent progression reasoning. Later phases, such as exploitation and post-compromise activity, provide stronger evidence and therefore require higher detection reliability. To balance missed detections and false alarms, individual phase predictions are stored as evidence but do not directly trigger an alarm. An alarm is raised only when multiple temporally consistent phases satisfy the predefined multi-step attack criteria. This preserves weak early-stage evidence while reducing the likelihood that an isolated false prediction results in an alarm, following a distinction between individual suspicious events and coordinated attack campaigns also found in HOLMES and ATLAS \cite{milajerdi2019holmes,alsaheel2021atlas}.

\subsection{Target Environment} \label{sec:target_environment} 
\name{} targets continuous multi-step attack detection in medium- to large-scale networks, where low inference latency, bounded state, and sustained flow processing are required. Because excessive false alarms can overwhelm security operators, the system is designed to raise an alarm only after multiple temporally consistent attack phases satisfy the predefined multi-step attack criteria. The system assumes access to live or captured network traffic processed by a network monitor capable of extracting structured flow records. To remain applicable when application traffic is encrypted, \name{} relies exclusively on packet-header and connection-level metadata, including protocol, endpoints, ports, direction, duration, and traffic-volume statistics. Payload inspection and host-only activity that produces no observable network evidence are outside the current scope.
\section{\name{}}
\label{sec:design_details}

\subsection{Overview}
\label{ssec:overview}
\name{} introduces a neuro-symbolic MSA detection system built using DeepProbLog \cite{manhaeve2018deepproblog,MANHAEVE2021103504}, combining neural attack-phase classification with probabilistic symbolic reasoning. 
Prior systems commonly reconstruct attack progression by correlating IDS alerts or causally related audit events into attack graphs and attack stories \cite{nadeem2021alert,milajerdi2019holmes,alsaheel2021atlas}. In contrast, \name{} integrates explicit attack-progression knowledge directly into flow-level inference, allowing previously observed phases to constrain subsequent attack-phase predictions.

\begin{figure*}
    \centering
    \includegraphics[width=\linewidth]{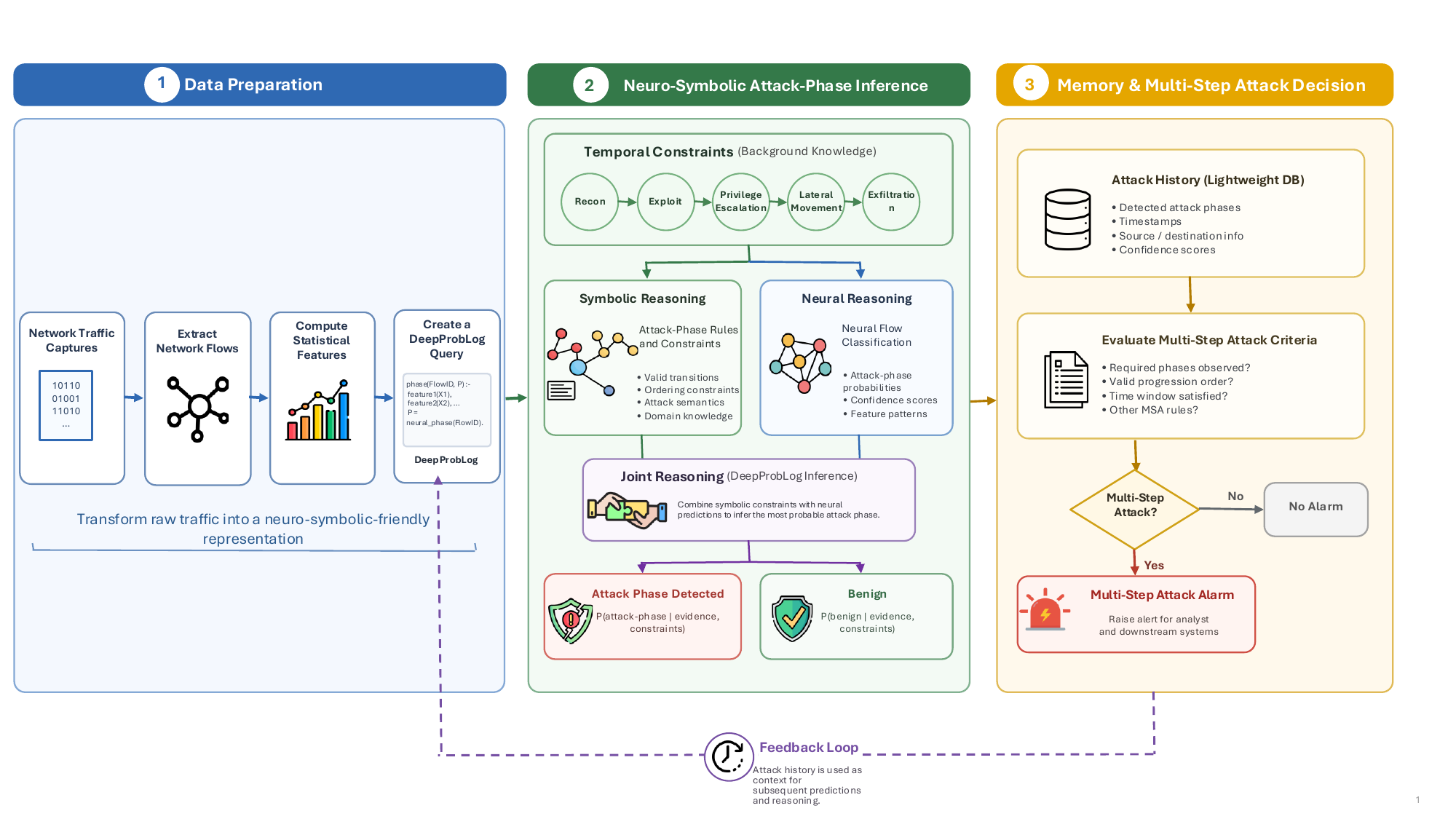}
    \caption{\name{} architecture. Network traffic is transformed into flow-level features and DeepProbLog queries, jointly evaluated through neural classification and symbolic temporal reasoning. Detected attack phases are stored as contextual history and evaluated against multi-step attack criteria to raise an alarm.}
    \label{fig:system_overview}
\end{figure*}

Figure~\ref{fig:system_overview} provides an overview of the proposed multi-step attack detection system. During data preparation, incoming network traffic captures are transformed into network flows, statistical features are computed, and DeepProbLog queries are constructed for neuro-symbolic attack-phase inference. During inference, both the neural and symbolic components reason about the most likely attack phase, after which a joint decision is made. If an attack phase is detected, it is stored in a lightweight database that maintains the attack history. This attack history is then used to provide contextual information for subsequent predictions. Finally, a multi-step attack alarm is raised if the accumulated attack history satisfies predefined MSA criteria.

Subsequently, we explain in detail each \name{} component, including:
\begin{inlinelist}
    \item the data preparation phase that transforms incoming raw traffic captures into a neuro-symbolic-friendly format that is used for inference (\Cref{ssec:data_preparation});
    \item the neuro-symbolic attack-phase inference that combines a neural component with temporal, symbolic reasoning (\Cref{ssec:nesy_inference}), and
    \item the multi-step attack detection decision (\Cref{ssec:memory_msa_decision}).
\end{inlinelist}

\subsection{Data Preparation}
\label{ssec:data_preparation}

The data preparation component is responsible for transforming incoming network traffic captures into a format that can be used for neuro-symbolic inference.

\ourparagraph{Network Flows.}
The first stage of the pipeline transforms raw network traffic captures into network flows. A network flow is defined as a grouping of packets sharing the same 5-tuple (source IP, destination IP, source port, destination port, and protocol) within a given time interval~\cite{NetworkFlow}.

This abstraction provides a condensed view of network activity, capturing communication patterns, volume statistics, and relationships between endpoints. In addition, flows are inherently sequential, allowing observations to be analyzed in the context of preceding and subsequent activity. This is particularly useful for detecting attack behaviors that only become apparent when considered across multiple events, making flow-based representations well suited for ML approaches that rely on temporal or relational structure.
Lastly, we design \name{} to operate on encrypted traffic and therefore exclusively extracting flow features from packet header fields. 

\ourparagraph{Statistical Features.}
Certain attack behaviors cannot be fully captured using individual network flows, as they emerge from patterns across multiple flows and over time. In particular, early-stage activities such as reconnaissance are often characterized by anomalous interaction patterns rather than properties of a single connection. For example, a single source IP probing multiple ports or multiple destination hosts may indicate scanning behavior \cite{MITRERECON}.
These patterns are inherently distributed and require a broader traffic context than that provided by a single flow. As a result, flow-level representations alone are insufficient to capture such global interaction patterns.

To address this limitation, and consistent with detection strategies in MITRE ATT\&CK \cite{MITREATTACK}, we compute both source-centric and destination-centric statistical features over sliding time windows. These features capture aggregated behavior across multiple flows and complement the flow-based representation. Importantly, they can still be derived from encrypted traffic, as they rely solely on packet header information such as timestamps, IP addresses, ports, and connection metadata.

\ourparagraph{Inference Query.}
\name{} operates over specific inference queries that are constructed from flow features -- including the statistical features -- and the attack history. Each query is passed to the neuro-symbolic inference component, which computes the probability $p$ over the possible query outcomes from $\{benign, phase_1, \ldots, phase_n\}$. 
For example, the probability of the current flow being classified as $phase_1$ is queried as:
\[
phase(C_{attack}, C_{flow}, Flow, phase_1),
\]
where $C_{attack}$ and $C_{flow}$ are the attack and flow contexts, respectively, and $Flow$ represents the flow features.

\paragraph{Attack Context.}
encodes which attack phases have been observed in the past. These are represented as binary attack-phase flags:

\[
C_{attack} = \{F_1, \ldots, F_n\}
\]

where $F_i = 1$ indicates that attack $phase_i$ has been already observed in the past (fed via a feed-back loop to the query construction as depicted in \Cref{fig:system_overview}), and $F_i = 0$ otherwise. 

\paragraph{Flow Context.}
contains the symbolic features used to evaluate whether a flow is consistent with known attack characteristics. These include protocol information, communication direction, and service-related attributes, which are commonly used in rule-based intrusion detection systems, such as Snort~\cite{roesch1999snort}, Suricata~\cite{oisf_suricata} or Zeek (formerly Bro)~\cite{paxson1999bro}.
Formally, a flow context is defined as:

\[
C_{flow} = (Protocol, Origin_{src}, Origin_{dst}, Port_{src}, Port_{dst}).
\]

\paragraph{Flow Features.}

The remaining component, $Flow$, represents the statistical features extracted from the flow and which are used as input to the neural network. We exclude timestamps, IP addresses, and port numbers from these features since they are highly environment-specific and can lead to poor generalization \cite{sarhan2022evaluating,sommer2010outside,cantone2024crossdataset}. In particular, attackers can easily change infrastructure, including IP addresses and port assignments, which may cause the neural model to learn patterns tied to a specific campaign rather than generalizable behavioral patterns. Instead, timestamps, IP addresses, and ports attributes are delegated to the symbolic component, which can explicitly encode and reason over them. This separation allows the neural network to focus on learning behavioral patterns that are more robust across different attack scenarios, while structural knowledge is handled through symbolic rules.

\subsection{Neuro-Symbolic Attack-Phase Inference}
\label{ssec:nesy_inference}

At a high level, an attack phase $phase_i$ is inferred when its temporal prerequisites are satisfied, the observed event is consistent with predefined attack knowledge, and the neural component indicates that the current flow is malicious. This combines temporal constraints, symbolic reasoning, and neural predictions into a unified inference process.
This is expressed in \name{} using DeepProbLog as follows:
{
\small
\[
\begin{aligned}
&\textcolor{blue}{\texttt{phase}}(C_{attack}, C_{flow}, Flow, phase_i) \; \leftarrow \\
&\quad \textcolor{purple}{\texttt{plausible\_attack\_phase}}(C_{attack} \wedge phase_i) \ \wedge \\
&\quad \textcolor{teal}{\texttt{attack\_phase\_rule}}(C_{flow}, phase_i) \ \wedge \\
&\quad \textcolor{orange}{\texttt{neural\_evidence}}(Flow, phase_i).
\end{aligned}
\]}
$\textcolor{purple}{\texttt{plausible\_attack\_phase}}$ is a logic predicate that defines the logically consistent attack phases which can be predicted given the previously seen past attack phases (see \Cref{sec:temporal_constraints}). 
This rule encodes, for example, that an exfiltration must follow a compromise, otherwise it is not achievable.
$\textcolor{teal}{\texttt{attack\_phase\_rule}}$ is a logic predicate defining the flow context characteristics that a traffic instance should match to be considered a plausible attack (see \Cref{sec:attack_phase_rules}).
This rule encode, for example, that a sadmind worm attempt must target port 111.
Lastly, the $\textcolor{orange}{\texttt{neural\_evidence}}$ represents the probability that the neural network assigns to the flow being suspicious.

On the other hand, a flow is considered benign if it does not match any of the defined attack phases. In other words, benign behavior is defined as the absence of evidence for all known attack phases and the corresponding rule is:
{
\small
\[
\begin{aligned}
&\textcolor{blue}{\texttt{phase}}(C_{attack}, C_{flow}, Flow, \texttt{benign}) \; \leftarrow \\
&\quad \neg\Big(
    \textcolor{purple}{\texttt{attack\_phase}}(phase_i) \ \wedge \\
&\qquad\quad
    \textcolor{blue}{\texttt{phase}}(C_{attack}, C_{flow}, Flow, phase_i)
\Big).
\end{aligned}
\]}
where $phase_i$ refers to any of the attack phases. 
In \name{}, we use the same DeepProbLog negation $\neg$ semantic expressing negation as negation-as-failure \cite{manhaeve2018deepproblog}.
Therefore, the probability of a negated goal is given by $1 - p$ and the probability of a benign flow can be expressed as:
\[
P(benign) = 1 - P\left(\bigvee_{i \in \{1,\dots,N\}} phase_i\right)
\]
In other words, a flow is considered benign with high probability when the probability of all attack phases is low.

\subsubsection{Temporal Constraints}\label{sec:temporal_constraints}
Due to the step-by-step nature of MSAs, an attack phase $phase_i$ can only occur if the preceding phase $phase_{i-1}$ has been observed. \name{} encodes this temporal structure in the symbolic component to restrict predictions to temporally plausible attack steps.

Consider a simplified setting with three ordered phases, where $phase_1$ initiates the attack, $phase_2$  depends on $phase_1$, and $phase_3$ depends on both preceding phases. $phase_1$ is therefore always considered valid, while $phase_2$  and $phase_3$  are conditioned on the presence of earlier phases.
This structure is encoded in \name{} using the following rules:

{\footnotesize
\[
\begin{aligned}
&\textcolor{purple}{\texttt{plausible\_attack\_phase}}
  (F_1, F_2, F_3, phase_1). \\
&\textcolor{purple}{\texttt{plausible\_attack\_phase}}
  (F_1, F_2, F_3, phase_2) \; \leftarrow  F_1 = 1. \\
&\textcolor{purple}{\texttt{plausible\_attack\_phase}}
  (F_1, F_2, F_3, phase_3) \; \leftarrow F_1 = 1 \wedge F_2 = 1. \\
\end{aligned}
\]
}

During grounding, only attack phases whose temporal constraints are satisfied will be retained. For example, given the flag configuration $(1,0,0)$, only $phase_1$ and $phase_2$ remain valid, while $phase_3$ is excluded due to the unmet prerequisite $F_2 = 1$.

\subsubsection{Attack Phase Rules}\label{sec:attack_phase_rules}
In addition to the temporal structure of MSA, each attack step exhibits characteristic patterns in terms of network behavior and protocol usage. \name{} encodes these characteristics as attack-phase rules in the symbolic component of the model. 
The use of attack-phase rules provides several benefits. First, the symbolic component enables reasoning over features such as IP addresses and ports which are not passed to the neural component. Additionally, the rule-based structure improves interpretability, as predictions can be traced back to explicit symbolic conditions.
For illustration, consider an attack $phase_i$ where an external attacker probes a victim for a vulnerable service running on port 123 using the TCP protocol. This can be expressed in \name{} using the following rule:

{\small
\[
\begin{aligned}
&\textcolor{teal}{\texttt{attack\_phase\_rule}}(
    Protocol, Origin_{src}, Origin_{dst}, \\
&\qquad Port_{dst}, phase_i) \; \leftarrow \\
&\quad Protocol = \texttt{tcp} \ \wedge Origin_{src} = \texttt{external} \ \wedge \\
&\quad Origin_{dst} = \texttt{internal} \ \wedge  Port_{dst} = 123.
\end{aligned}
\]
}
Using such rule, \name{} will not predict an attack if the port 123 is not being probed by the attacker.

\subsection{Memory and Multi-Step Attack Decision}
\label{ssec:memory_msa_decision}

The decision and memory components decouple higher-level operational logic from the neuro-symbolic inference process. These decisions, such as when to trigger an MSA alert or how long to retain predictions in memory, do not require neuro-symbolic reasoning. On the contrary, both components are designed to be configurable by security operators to support different deployment environments.

\ourparagraph{Attack Memory.}
As the neuro-symbolic inference component itself is stateless, \name{} maintains a lightweight attack memory to preserve contextual information across flows. The memory stores previously observed attack phases for each tracked attack history state $h$, defined by the IP addresses involved in the attack. This ensures that retrieved attack-phase flags remain relevant to the specific network flow being analyzed. For example, assume an attacker $IP_{A1}$ first performs a scan and later exploits a vulnerability on the target $IP_{T1}$. In this case, several attack-phase flags will be set for the context associated with $(IP_{A1}, IP_{T1})$. However, these flags should not influence inference for a different attacker $IP_{A2}$ interacting with the network. By maintaining separate history states, the system can simultaneously track multiple attack campaigns while preventing unrelated attack histories from affecting predictions.

When an attack phase is detected, a new memory entry is created:
{\small
\[
\begin{aligned}
\big<
h :\;& (time_1, \texttt{attack\_phase}_1, probability_1)
\big>
\end{aligned}
\]
}
As additional attack phases are observed, the memory associated with the context is updated accordingly:
{\small
\[
\begin{aligned}
\big<
h :\;& (time_1, \texttt{attack\_phase}_1, probability_1), \ldots, \\
     & (time_i, \texttt{attack\_phase}_i, probability_i)
\big>,
\end{aligned}
\]
}
where, $time_i$ denotes the start time of the observed attack phase, $attack\_phase_i$ represents the detected phase, and $probability_i$ corresponds to the confidence assigned by the neuro-symbolic model.

When processing a new flow, the system queries the memory using the IPs of the current flow. If a matching attack history state exists, the corresponding attack context $C_{attack}$ is retrieved and passed to the neuro-symbolic component to guide inference. History states are maintained for a configurable retention period and are removed once they exceed the retention window without further activity.  
This memory component enables stateful detection of stealthy multi-step attacks that unfold gradually over extended periods of time. Additionally, it provides information about attack progression, which is useful for security analysts when deciding whether to initiate further investigation.

\ourparagraph{MSA Alarm Decision.}
The final alarm decision is based on the accumulation of inferred attack phases within a given attack history. Rather than triggering alarms based on isolated malicious predictions, the system considers the temporal progression of phases as evidence of an ongoing MSA.
This enables the incorporation of early attack phases such as reconnaissance, which are often noisy but important for early-stage detection. Individually, such phases may produce a large number of false positives. However, by requiring multiple related attack phases to occur within the same history, many of these isolated false positives can be filtered out before an alarm is raised.

An alarm is triggered when a history $h$ satisfies a configurable detection condition based on the set of previously observed attack phases and their associated probabilities. By default, the condition requires the observation of at least two distinct attack phases within the same history. This reduces the likelihood that noisy or isolated predictions result in unnecessary alerts.
This alerting mechanism can be adapted to different operational environments by adjusting detection thresholds and phase requirements without modifying the underlying neuro-symbolic inference model.
\section{Implementation}
\label{sec:implementation}

We implement \name{} in Python using PyTorch~\cite{paszke2019pytorch} and
DeepProbLog~\cite{MANHAEVE2021103504}.
%
%
The neural component of \name{} is a Long Short-Term Memory (LSTM) classifier, consisting of a single recurrent layer with 64 hidden units, followed by a fully connected layer with 32 neurons using the ReLu activation function and a final softmax layer. 
The architecture is intentionally kept simple to ensure that performance improvements can be more reliably attributed to the integration of symbolic reasoning rather than model complexity.
Additionally, we adopt a binary classification setup, as different attack phases may exhibit similar low-level patterns and are therefore difficult to distinguish directly. 
The neural network is incorporated into the DeepProbLog model as a neural predicate, which outputs probabilities over the classes \textit{benign} and \textit{malicious}:

{\small
\[
\begin{aligned}
&\texttt{nn}(\texttt{lstm}_{binary}, [Flow], Event,
[\texttt{benign}, \texttt{malicious}]) \mathbin{::} \\
&\quad \textcolor{orange}{\texttt{neural\_evidence}}(Flow, Event).
\end{aligned}
\]
}

Temporal constraints in MSA detection can be implemented with varying degrees of strictness. In particular, one may allow flexibility in the observed attack sequence -- e.g., regression to earlier phases -- or enforce a strictly forward progression through the attack stages.
This choice directly affects the trade-off between false alarms and missed detections.
More permissive rules reduce the number of missed attacks but may increase spurious transitions, while stricter rules reduce ambiguity at the cost of reduced flexibility.
\name{} adopts a strict temporal ordering where an attack phase $p_i$ is only considered plausible if its prerequisite phase $p_{i-1}$ has been observed, and none of the subsequent phases $p_{i+1}$ have yet occurred within the same context. Since the model operates on fully observed datasets with ground-truth attack progressions, the risk of missing earlier phases due to this constraint is not a concern in our setting.
Formally, we encode:

{\small
\[
\begin{aligned}
&\textcolor{purple}{\texttt{plausible\_attack\_phase}}
(F_1, F_2, F_3, phase_1) \; \leftarrow \\
&\quad F_1 = 0 \ \wedge \; F_2 = 0 \ \wedge \; F_3 = 0. \\[0.4em]
&\textcolor{purple}{\texttt{plausible\_attack\_phase}}
(F_1, F_2, F_3, phase_2) \; \leftarrow \\
&\quad F_1 = 1  \ \wedge \; F_2 = 0 \ \wedge \; F_3 = 0. \\[0.4em]
&\textcolor{purple}{\texttt{plausible\_attack\_phase}}
(F_1, F_2, F_3, phase_3) \; \leftarrow \\
&\quad F_1 = 1 \ \wedge \; F_2 = 1 \ \wedge \; F_3 = 0.
\end{aligned}
\]
}

The temporal rules restrict inference to only two possible outcomes at a given step, i.e., the next attack phase or benign traffic. Consequently, the grounded program simplifies to

\[
P(benign) = 1-P(NextAttackPhase)
\]


\name{} implements attack-phase rules encoding domain knowledge about the network characteristics derived from documented attack behavior, vulnerability descriptions, expert knowledge, or established IDS rulesets such as the Snort Community Rules \cite{SNORTCOMMUNITYRULES,roesch1999snort}. Because \name{} does not inspect packet payloads, payload-dependent conditions are omitted and only flow-observable properties are retained. Satisfying a rule does not independently confirm an attack phase. Instead, the rule provides symbolic evidence that is combined with the neural classifier's output during \name{} inference. The following examples illustrate distinct attack-phase rules implemented in \name{}. 

\paragraph{Network Scanning.} An inbound ICMP scan is characterized by ICMP traffic originating outside the monitored network and targeting internal hosts: 
\[ \small
\begin{aligned} 
&
scanning\_rule(Protocol, Origin_{src}, Origin_{dst}) \; \leftarrow \\ 
& Protocol = icmp \ \wedge Origin_{src} = external \ \wedge \\ 
& Origin_{dst} = internal. 
\end{aligned} 
\] 

\paragraph{Service Probing.} The service-probing rule abstracts the header-level conditions of a Snort signature for \texttt{sadmind} probing associated with CVE-1999-0977 \cite{CVE19990977,SNORTCOMMUNITYRULES}. Payload-matching conditions are excluded: 
\[ 
\small
\begin{aligned} 
& probing\_rule(Protocol, Origin_{src}, Origin_{dst}) \; \leftarrow \\
& Protocol = udp \ \wedge Origin_{src} = external \ \wedge \\
& Origin_{dst} = internal.
\end{aligned}
\] 

\paragraph{DNS-Based Data Exfiltration.} DNS-based exfiltration is represented by internal UDP communication directed toward a DNS service: 
\[ 
\small
\begin{aligned} 
& exfiltration\_rule(Protocol, Origin_{src}, Origin_{dst}, Port_{dst}) \; \leftarrow \\
& Protocol = udp \ \wedge Origin_{src} = internal \ \wedge \\
& Origin_{dst} = internal \ \wedge Port_{dst} = 53.
\end{aligned}
\] 
Because these conditions also describe legitimate DNS traffic, the rule only establishes compatibility with the expected communication channel and complementary statistical evidence is supplied by the neural classifier. 

\paragraph{Rule Encoding.} Categorical attributes are numerically encoded before inference. Protocols use their assigned IP protocol numbers \cite{IPPROTOCOLNUMBERS}, while endpoint locations are represented as binary indicators, with \(1\) denoting internal and \(0\) external. For example, the probing rule is encoded as: 
\[
\small
\begin{aligned}
& probing\_rule(Protocol, Origin_{src}, Origin_{dst}) \; \leftarrow \\
& Protocol = 17 \ \wedge Origin_{src} = 0 \ \wedge Origin_{dst} = 1. \end{aligned} 
\] 
where \(17\) represents UDP. This encoding provides a uniform interface between flow-level data and symbolic reasoning.
\section{Evaluation}
\label{sec:evaluation}

This section depicts the results of \name{} evaluation.
First, we describe our experimental settings, including datasets, evaluation metrics and baselines (\Cref{ssec:experimental_setup}).
Then, we elaborate the effectiveness of \name{} in different scenarios (\Cref{ssec:effectiveness}) and conduct a missed attack steps analysis and assess the tolerance of the temporal constraints (\Cref{ssec:missed_attack_steps}). 
Lastly, we analyze performance overhead \name{} performance, including latncy, throughput and peak memory usage (\Cref{ssec:performance_overhead}).

\subsection{Experimental Setup}
\label{ssec:experimental_setup}
%
\name{} is evaluated using two publicly available MSA datasets: 
DARPA 2000~\cite{DARPA2000} and AIT Log Data Set V2 (AIT-LDS V2)~\cite{landauer2021have,landauer2023maintainable}.
Both datasets contain network traffic packet captures (PCAPs) and labeled attack phases. The two datasets are chosen to provide complementary perspectives, combining a widely used benchmark with a more recent and realistic traffic environment. The datasets were divided into training and testing sets using a stratified 70/30 split, which ensured that samples from all attack phases were represented in both subsets. More details related to the datasets can be found in Appendix~\ref{secapp:dataset_details}.
For both datasets, packet captures are transformed into Zeek flows, aligned with the available attack annotations, and converted into temporal neural inputs and inference queries. Full preprocessing details and the resulting class distributions are provided in Appendix~\ref{app:data_preparation}.


\ourparagraph{Metrics.}
%
%
\name{} performs a multi-class classification, and therefore several standard classification metrics are recorded, including accuracy, F1-scores (macro and weighted), false alarm rate (FAR), and detection rate (DR). In addition, per-phase F1-scores are computed. Accuracy and F1-scores are evaluated over all classes, both benign traffic and attack phases. The false alarm rate is computed as FAR = \(\frac{\# \text{ benign predicted as attack}}{\# \text{ true benign samples}}\), while the detection rate is computed as DR = \(\frac{\# \text{ correctly detected attacks}}{\# \text{ true attack samples}}\).


%
To capture the \name{}'s ability to learn the temporal dependencies of MSA, we define two custom temporal metrics: causal violation rate (CVR) and regression violation rate (RVR). 
Such metrics represent two different types of temporal inconsistencies. A causal violation occurs when the model predicts attack phase \(p_{i+1}\) before phase \(p_i\) has been observed. A regression violation occurs when the model has already progressed to phase \(p_i\) but subsequently predicts an earlier attack phase. 
The causal violation rate is defined as CVR = \(\frac{\# \text{ causal violations }}{\# \text{ total violations }}\), while the regression violation rate is defined as RVR = \(\frac{\# \text{ regression violations}}{\# \text{ total violations }}\).
Causal violations are considered more severe than regression violations. From a logical perspective, causal violations should not occur, as each attack step depends on the prior occurrence of prerequisite steps. Thus, CVR pinpoints instances where the detection model fails to take into account the proper attack temporal logic. Regression violations, on the other hand, may occur in practice if an attacker revisits previous attack stages. Nevertheless, they still represent temporal inconsistencies which should be possible to minimize through the incorporation of logic.

\ourparagraph{Baselines.}
\name{} is compared to both nerual and symbolic-based baselines, and state-of-the-art systems including Kitsune \cite{mirsky2018kitsune}.
We implemented two neural baselines: a multiclass LSTM and an ensemble LSTM. Both models use the same underlying architecture as the neural component of \name{}. However, rather than performing binary classification, the baseline models directly predict the attack step through a multiclass output layer. The ensemble architecture is similar to the multiclass model but replaces the shared LSTM with multiple LSTMs (one per attack phase), whose outputs are combined through a linear prediction layer.
The symbolic baselines are the \name{} logic programs with the neural component removed. This isolates the contribution of the logical reasoning component and provides insight into the extent to which performance can be attributed to the logic alone. Comparing these baselines with the full model allows us to assess whether the neural network provides meaningful predictive capability beyond that achieved by the logical rules.

For \name{}, we evaluate two variants: 
\begin{inlinelist}
    \item an end-to-end (E2E) training version -- where the neural and symbolic components are optimized jointly from scratch --, and
    \item its neurally pretrained (Pretrained) counterpart---where the neural component is first trained independently on the training data and subsequently integrated into the \name{} logic program before being fine-tuned together with the symbolic component. 
\end{inlinelist}

Both the baseline models and the \name{} models are trained for a maximum of 50 epochs.
The neural baselines are evaluated using the dataset-specific hyperparameter configurations selected through the ablation study in Appendix~\ref{sec:neural_ablation_study}. Specifically, we use the base feature set and a learning rate of \(10^{-3}\) for both datasets, with temporal windows of \(100\) flows for AIT-LDSv2 and \(10\) flows for DARPA 2000.

\subsection{Effectiveness}
\label{ssec:effectiveness}

\ourparagraph{Symbolic Knowledge Study.}
This experiment investigates whether incorporating symbolic knowledge can help the model reliably detect individual attack steps and enforce temporally valid attack progressions. We evaluate two \name{} variants: 
\begin{inlinelist}
    \item one that incorporates temporal constraints only (T), and
    \item another that incorporates both temporal constraints and attack-phase rules (T+R).
\end{inlinelist}
The \name{} models are compared against the neural baselines, symbolic baselines, and Kitsune~\cite{mirsky2018kitsune}. \Cref{tab:combined_temporal_consistency} depicts the results.
On AIT-LDSv2, incorporating both temporal constraints and attack-phase rules (T+R) yields the strongest results, increasing macro F1 from \(0.878\) to \(0.944\) for the pretrained model and reducing its RVR from \(0.152\) to \(0.002\). On DARPA, the end-to-end T+R model achieves the best \name{} macro F1-score (\(0.940\)) and perfect temporal consistency (\(\mathrm{CVR}=\mathrm{RVR}=0\)), while the neural baselines exhibit substantially more progression violations. Although Kitsune obtains the highest macro F1-score on DARPA (\(0.984\)), its FAR of \(0.03719\) is considerably higher than that of the end-to-end T+R model (\(0.00015\)), and, as a single-event anomaly detector, it does not provide attack-progression consistency metrics.

Neither training strategy consistently dominates across datasets. Pretraining performs best on AIT-LDSv2, while end-to-end optimization substantially improves the T+R model on DARPA. These results suggest that joint training is beneficial when the symbolic rules are strongly aligned with the observed attack behavior, whereas pretraining can be more robust when symbolic knowledge is incomplete or only weakly discriminative.
\begin{table*}[t]
\centering
\fontsize{8}{9.5}\selectfont   
\caption{Performance and temporal consistency on AIT-LDS V2 and DARPA datasets. Results for AIT-LDS V2 are averaged over five runs, except \name{}\textsubscript{Pretrained} T, which is averaged over three runs. Results for DARPA are averaged over three runs. Standard deviations are reported only for macro F1. Models are ranked by macro F1-score within each dataset. Best results for CVR and RVR within each dataset are highlighted in bold.}
\label{tab:combined_temporal_consistency}
\begin{tabular}{lccccccc}
\toprule
Model & Macro F1 ($\pm$ std) & Weighted F1 & Accuracy & DR & FAR & CVR & RVR \\
\midrule
\multicolumn{8}{l}{\textit{AIT-LDS V2}} \\
\name{}\textsubscript{Pretrained} T+R & 0.944 $\pm$ 0.024 & 0.999 & 0.999 & 0.990 & 0.0003 & \textbf{0.000} & \textbf{0.002} \\
\name{}\textsubscript{E2E} T+R & 0.940 $\pm$ 0.010 & 0.999 & 0.999 & 0.991 & 0.0005 & \textbf{0.000} & 0.014 \\
LSTM\textsubscript{Multiclass} & 0.880 $\pm$ 0.015 & 0.998 & 0.998 & 0.988 & 0.0018 & 0.019 & 0.021 \\
\name{}\textsubscript{Pretrained} T & 0.878 $\pm$ 0.020 & 0.998 & 0.998 & 0.983 & 0.0008 & \textbf{0.000} & 0.152 \\
LSTM\textsubscript{Ensemble} & 0.858 $\pm$ 0.024 & 0.998 & 0.998 & 0.991 & 0.0019 & 0.027 & 0.017 \\
\name{}\textsubscript{E2E} T & 0.842 $\pm$ 0.013 & 0.998 & 0.998 & 0.981 & 0.0008 & \textbf{0.000} & 0.180 \\
Kitsune~\cite{mirsky2018kitsune} & 0.645 & 0.938 & 0.947 & 0.231 & 0.0125 & -- & -- \\
Logic\textsubscript{Baseline} T+R & 0.631 $\pm$ 0.038 & 0.917 & 0.925 & 0.714 & 0.0627 & \textbf{0.000} & 0.047 \\
Logic\textsubscript{Baseline} T & 0.092 $\pm$ 0.000 & 0.100 & 0.053 & 1.000 & 1.000 & \textbf{0.000} & 0.200 \\
\midrule
\multicolumn{8}{l}{\textit{DARPA}} \\
Kitsune~\cite{mirsky2018kitsune} & 0.984 & 0.984 & 0.984 & 0.999 & 0.03719 & -- & -- \\
\name{}\textsubscript{E2E} T+R & 0.940 $\pm$ 0.014 & 0.999 & 0.999 & 0.999 & 0.00015 & \textbf{0.000} & \textbf{0.000} \\
\name{}\textsubscript{Pretrained} T+R & 0.892 $\pm$ 0.006 & 0.999 & 0.999 & 0.999 & 0.00009 & \textbf{0.000} & \textbf{0.000} \\
Logic\textsubscript{Baseline} T+R & 0.891 $\pm$ 0.079 & 0.963 & 0.962 & 0.999 & 0.09021 & \textbf{0.000} & 0.001 \\
\name{}\textsubscript{Pretrained} T & 0.872 $\pm$ 0.013 & 0.999 & 0.999 & 0.999 & 0.00021 & \textbf{0.000} & \textbf{0.000} \\
\name{}\textsubscript{E2E} T & 0.839 $\pm$ 0.019 & 0.999 & 0.999 & 0.999 & 0.00015 & \textbf{0.000} & 0.014 \\
LSTM\textsubscript{Multiclass} & 0.827 $\pm$ 0.016 & 0.999 & 0.999 & 0.999 & 0.00030 & 0.027 & 0.081 \\
LSTM\textsubscript{Ensemble} & 0.820 $\pm$ 0.014 & 0.999 & 0.999 & 0.999 & 0.00032 & 0.053 & 0.092 \\
Logic\textsubscript{Baseline} T & 0.558 $\pm$ 0.000 & 0.655 & 0.585 & 1.000 & 0.99834 & \textbf{0.000} & 0.144 \\
\bottomrule
\end{tabular}
\end{table*}

\begin{figure*}[!t]
\centering
\captionsetup[subfigure]{font=footnotesize, skip=2pt}
\setlength{\abovecaptionskip}{3pt}
\setlength{\belowcaptionskip}{-4pt}

\begin{subfigure}{0.32\linewidth}
    \centering
    \includegraphics[width=\linewidth]
    {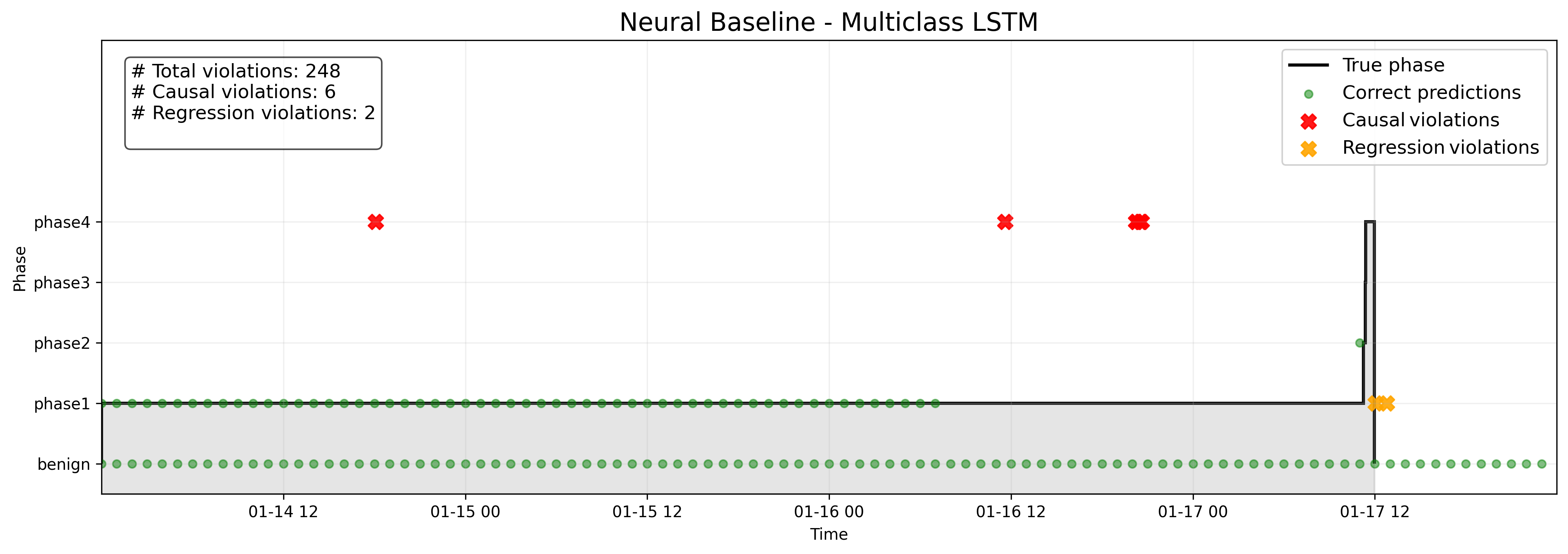}
    \caption{LSTM$_{Multiclass}$}
    \label{fig:lstm_multiclass}
\end{subfigure}
\hfill
\begin{subfigure}{0.32\linewidth}
    \centering
    \includegraphics[width=\linewidth]
    {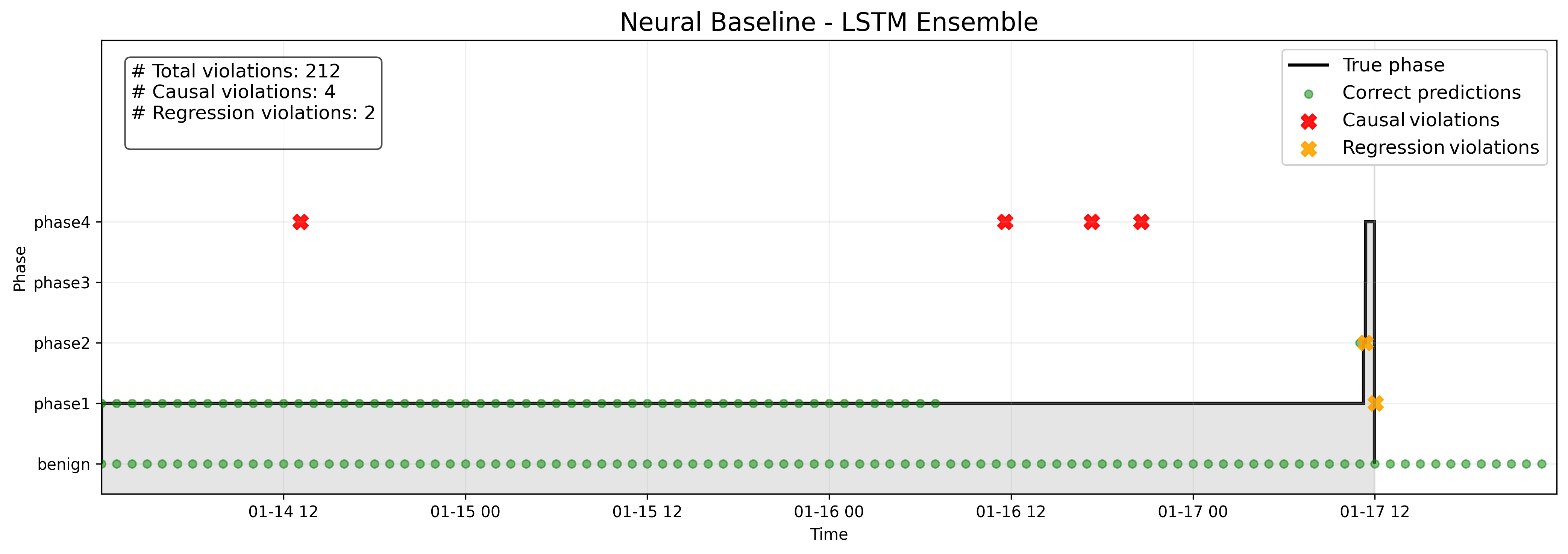}
    \caption{LSTM$_{Ensemble}$}
    \label{fig:lstm_ensamble}
\end{subfigure}
\hfill
\begin{subfigure}{0.32\linewidth}
    \centering
    \includegraphics[width=\linewidth]
    {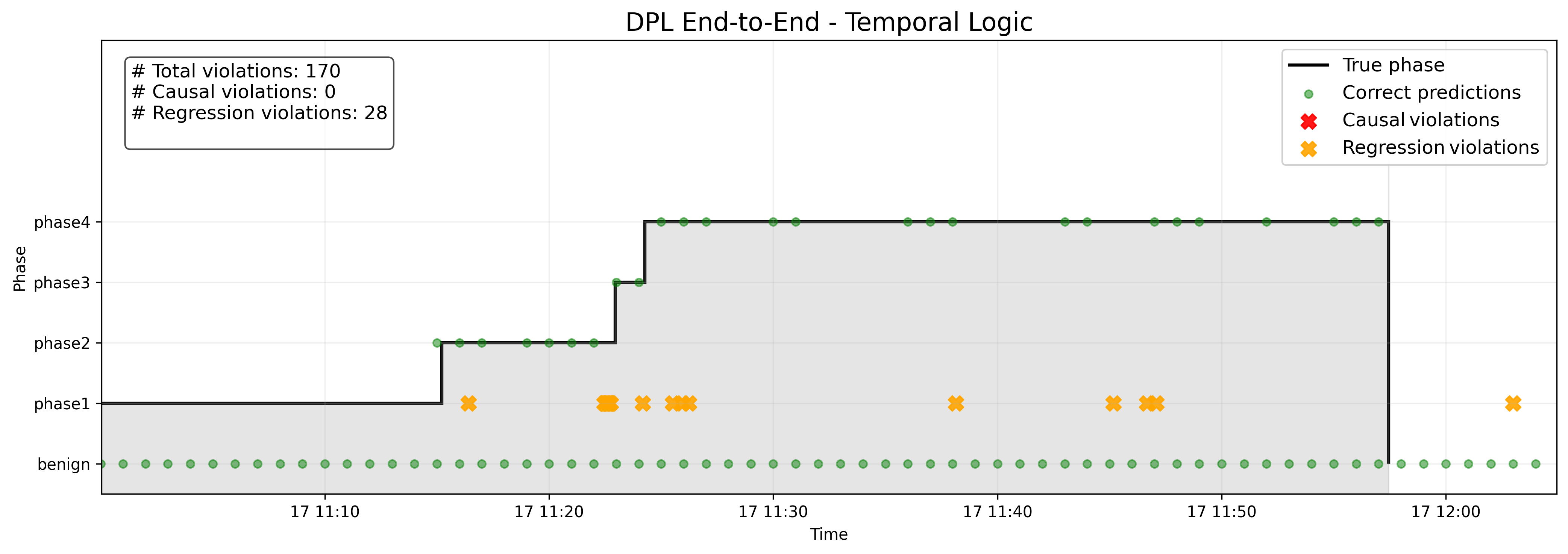}
    \caption{\name{}$_{E2E}$ (T)}
    \label{fig:deepchronos_e2e_t}
\end{subfigure}

\vspace{0.3em}

\begin{subfigure}{0.32\linewidth}
    \centering
    \includegraphics[width=\linewidth]
    {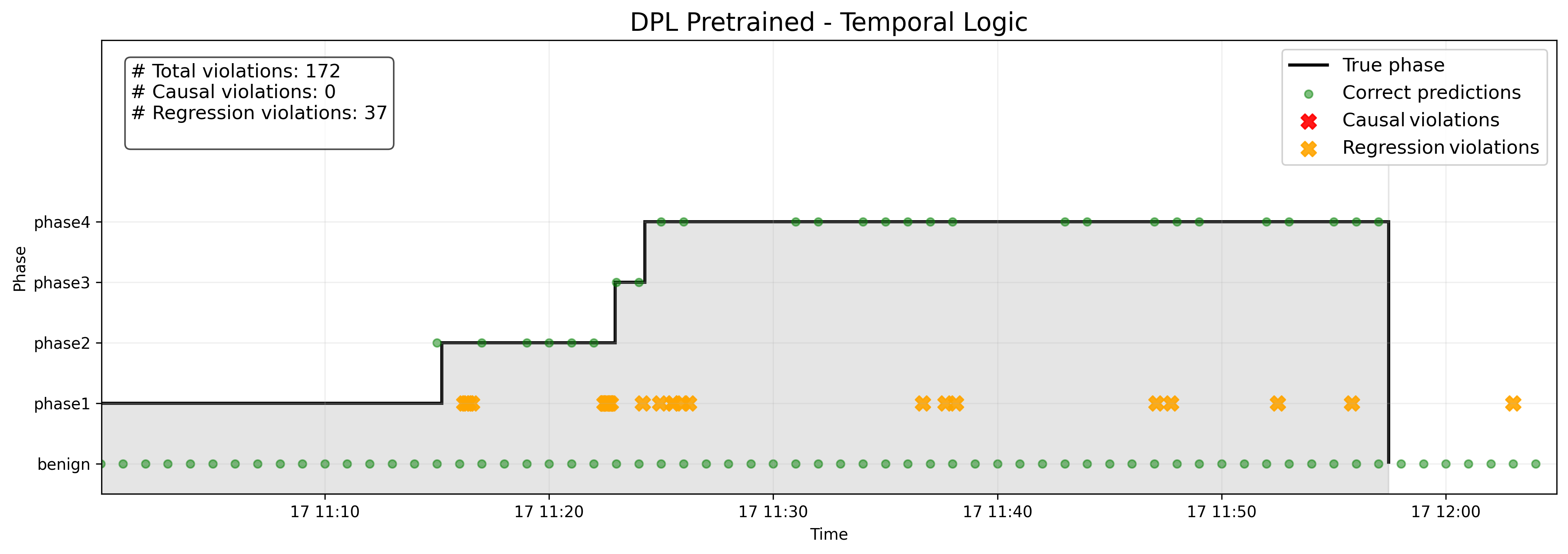}
    \caption{\name{}$_{Pretrained}$ (T)}
    \label{fig:deepchronos_pretrained_t}
\end{subfigure}
\hfill
\begin{subfigure}{0.32\linewidth}
    \centering
    \includegraphics[width=\linewidth]
    {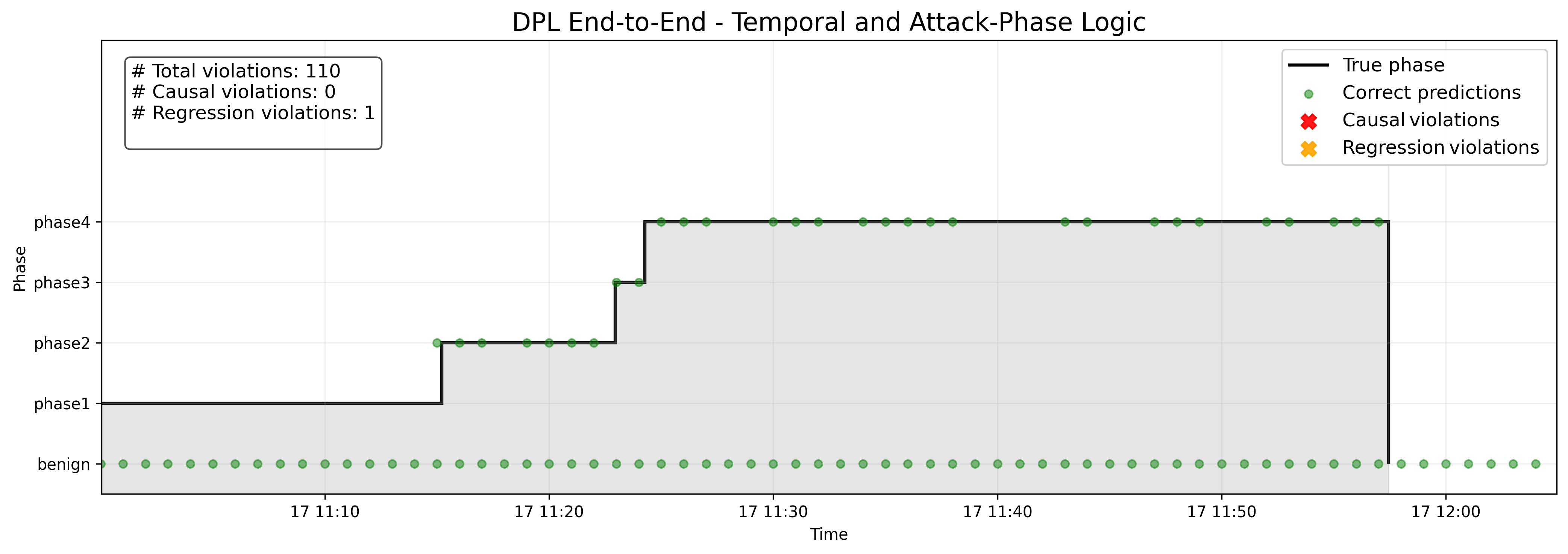}
    \caption{\name{}$_{E2E}$ (T+R)}
    \label{fig:deepchronos_e2e_tr}
\end{subfigure}
\hfill
\begin{subfigure}{0.32\linewidth}
    \centering
    \includegraphics[width=\linewidth]
    {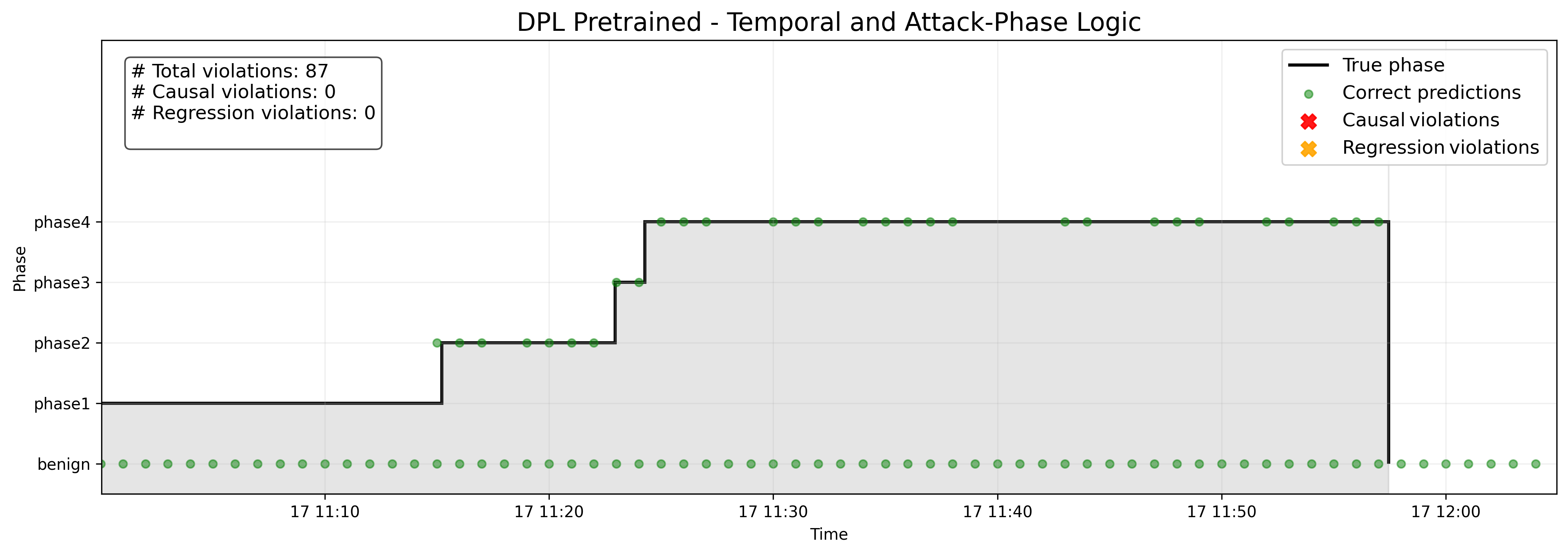}
    \caption{\name{}$_{Pretrained}$ (T+R)}
    \label{fig:deepchronos_pretrained_tr}
\end{subfigure}

\caption{
Temporal consistency comparison across neural and neuro-symbolic models on AIT-LDS V2.
}
\label{fig:ait_temporal_plot}
\end{figure*}

\begin{figure*}[!t]
\centering
\captionsetup[subfigure]{font=footnotesize, skip=2pt}
\setlength{\abovecaptionskip}{3pt}
\setlength{\belowcaptionskip}{-4pt}

\begin{subfigure}{0.32\linewidth}
    \centering
    \includegraphics[width=\linewidth]
    {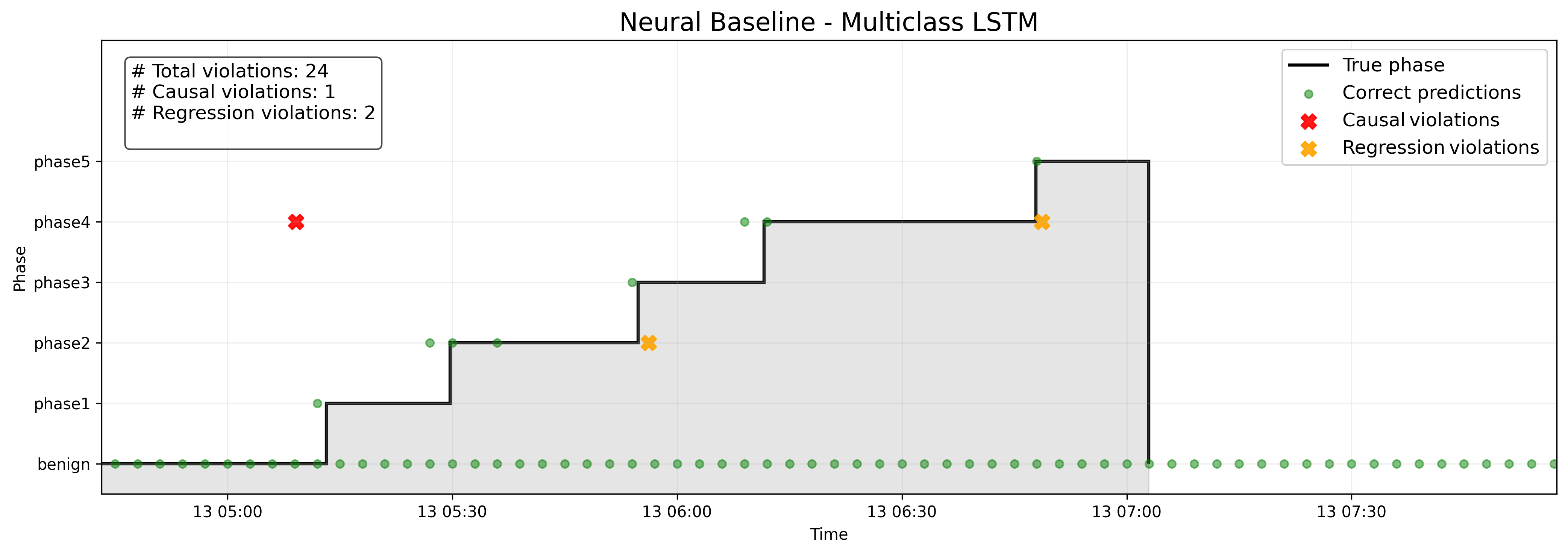}
    \caption{LSTM$_{Multiclass}$}
\end{subfigure}
\hfill
\begin{subfigure}{0.32\linewidth}
    \centering
    \includegraphics[width=\linewidth]
    {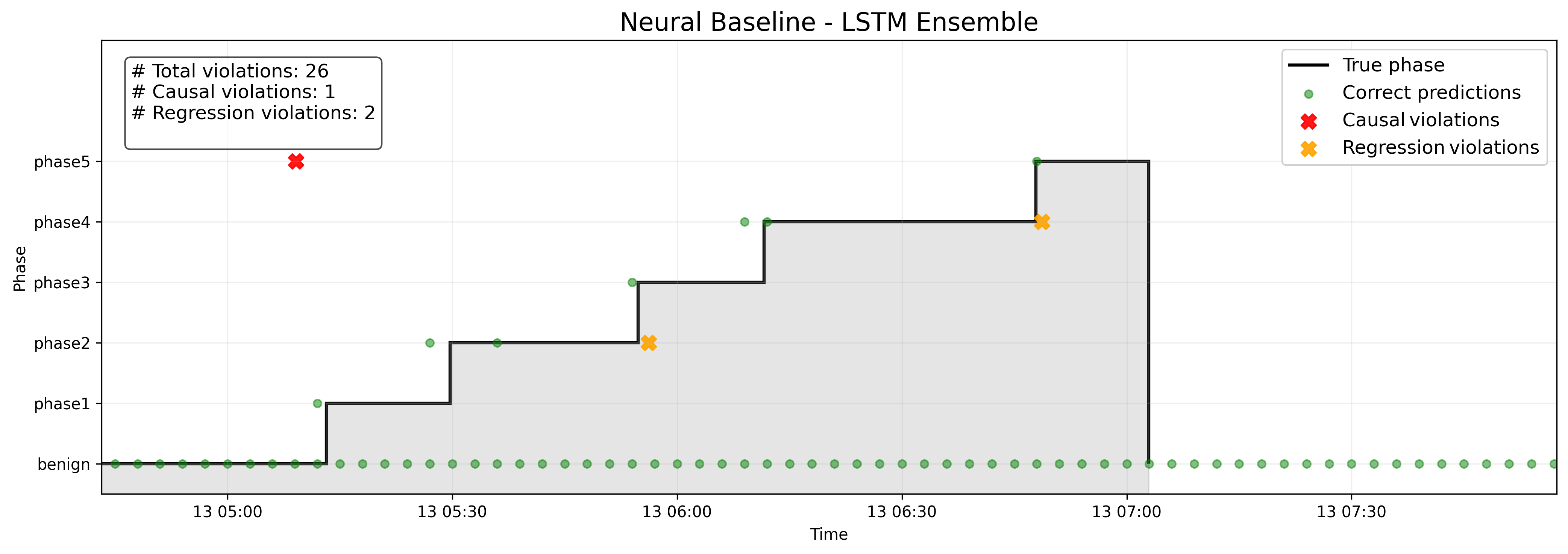}
    \caption{LSTM$_{Ensemble}$}
\end{subfigure}
\hfill
\begin{subfigure}{0.32\linewidth}
    \centering
    \includegraphics[width=\linewidth]
    {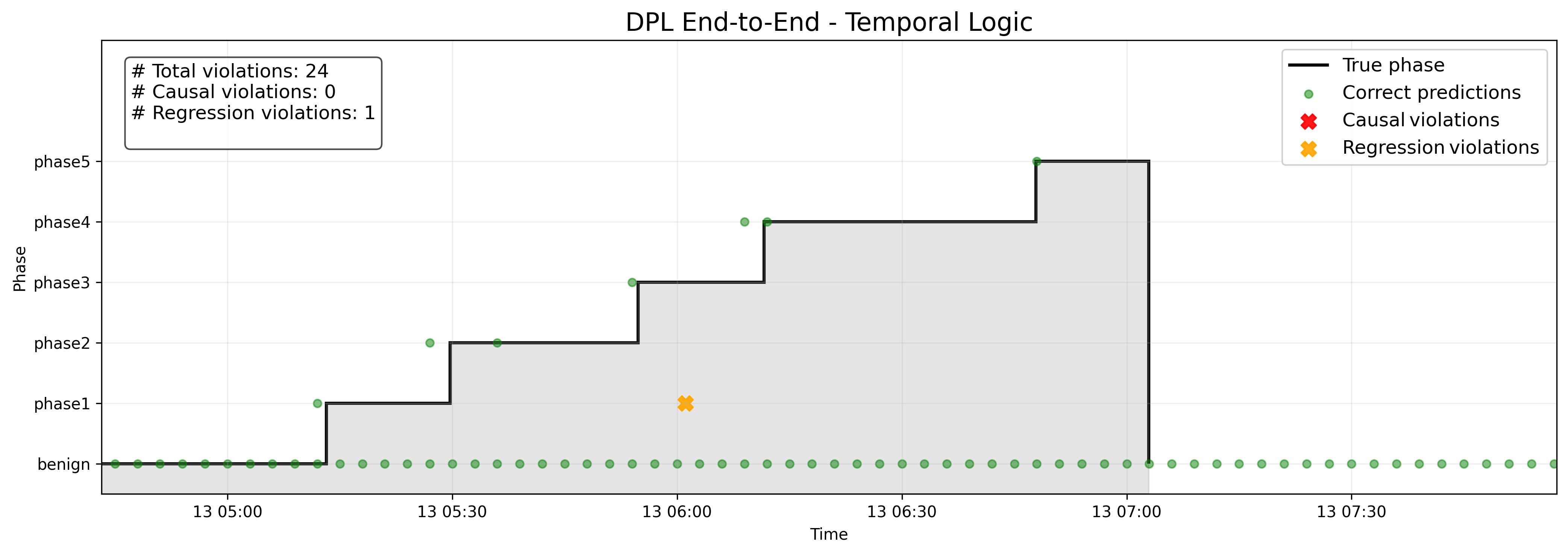}
    \caption{\name{}$_{E2E}$ (T)}
\end{subfigure}

\vspace{0.3em}

\begin{subfigure}{0.32\linewidth}
    \centering
    \includegraphics[width=\linewidth]
    {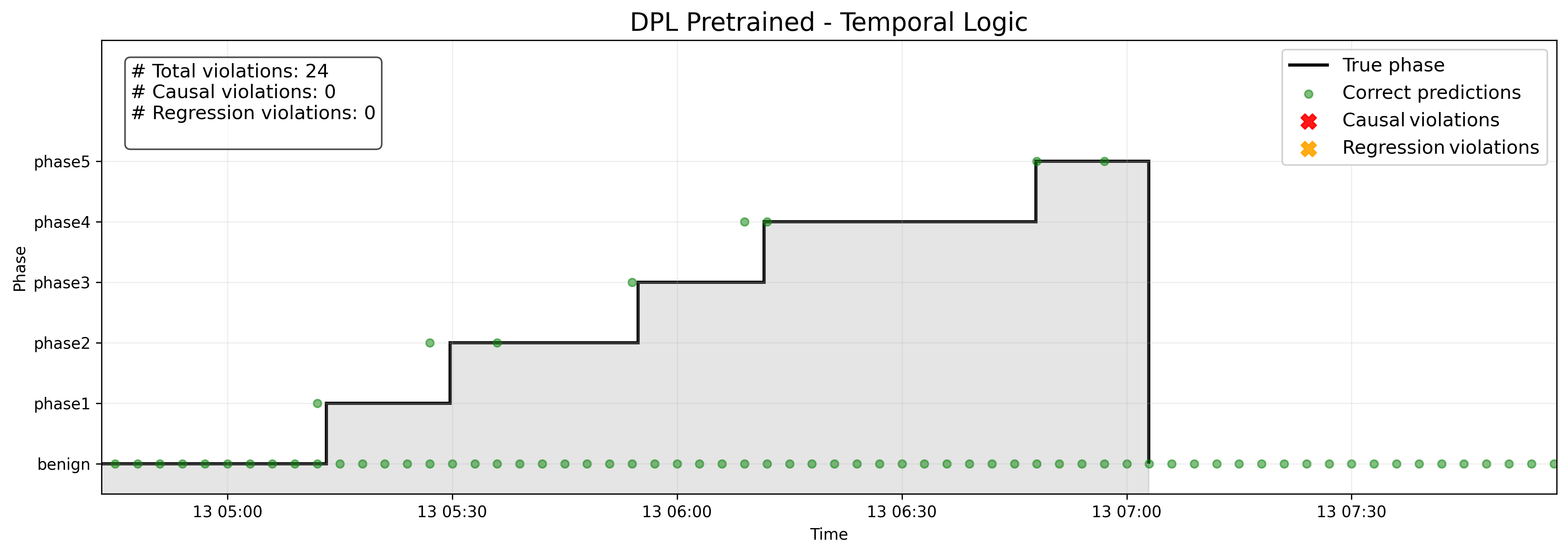}
    \caption{\name{}$_{Pretrained}$ (T)}
\end{subfigure}
\hfill
\begin{subfigure}{0.32\linewidth}
    \centering
    \includegraphics[width=\linewidth]
    {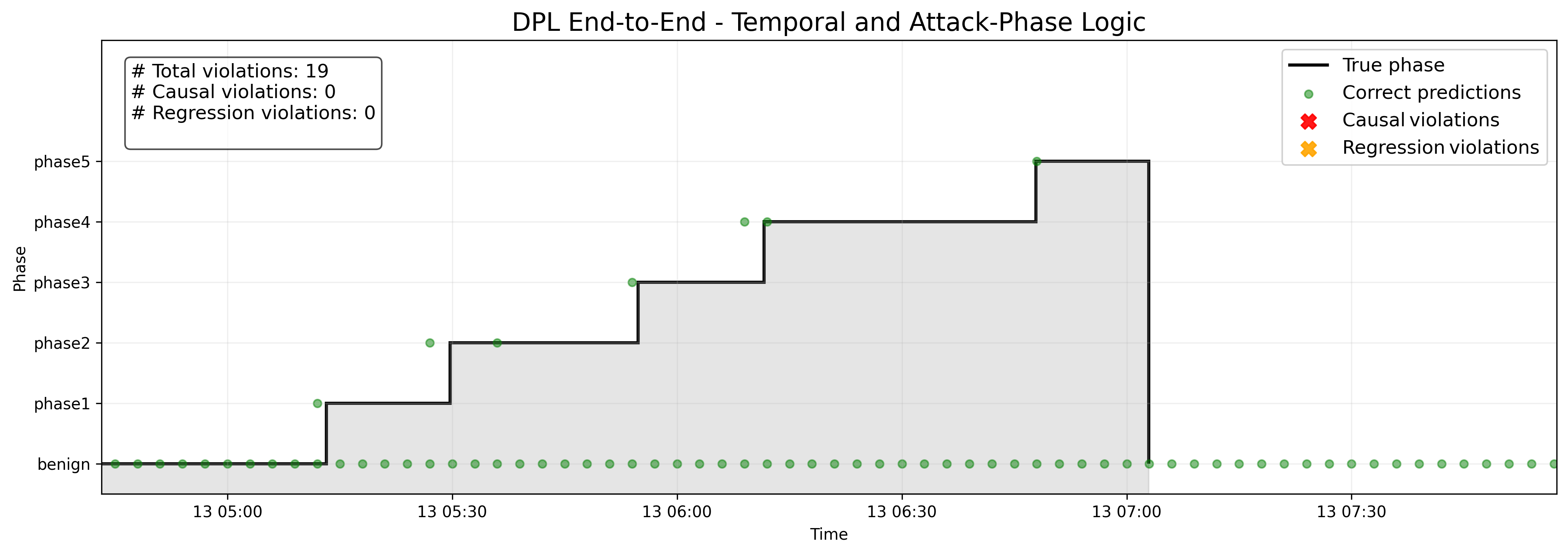}
    \caption{\name{}$_{E2E}$ (T+R)}
\end{subfigure}
\hfill
\begin{subfigure}{0.32\linewidth}
    \centering
    \includegraphics[width=\linewidth]
    {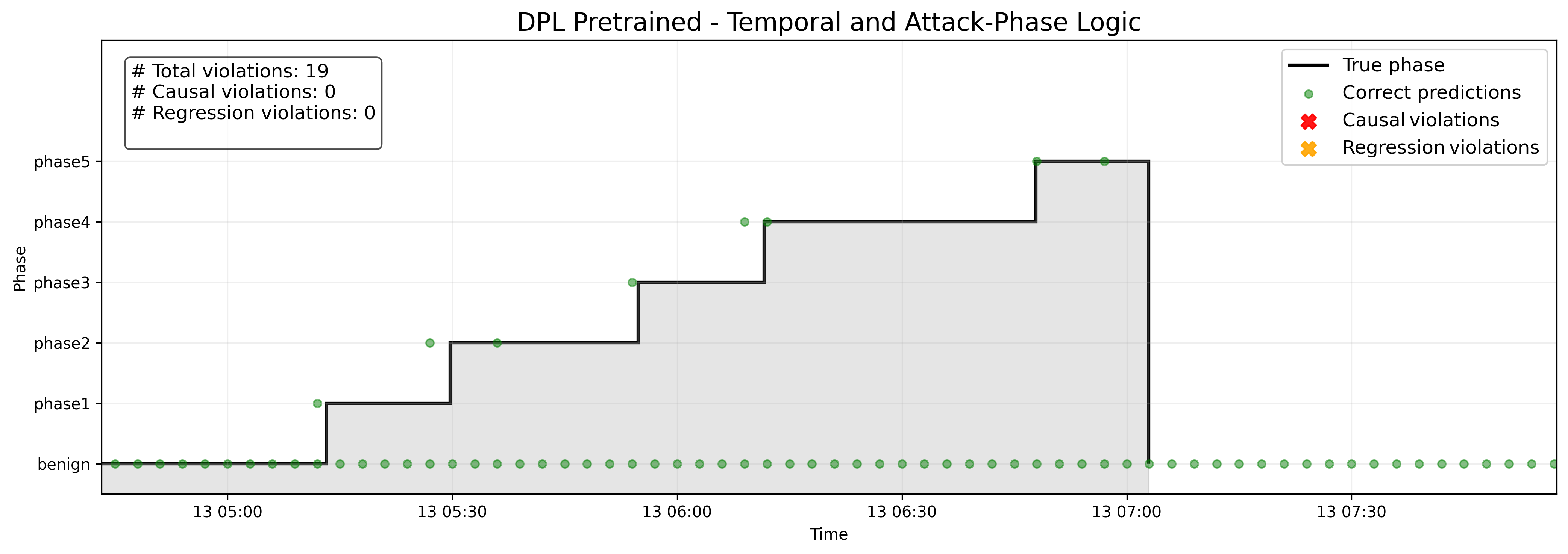}
    \caption{\name{}$_{Pretrained}$ (T+R)}
\end{subfigure}

\caption{
Temporal consistency comparison across neural and neuro-symbolic models on DARPA 2000.
}
\label{fig:darpa_temporal_plot}
\end{figure*}

\ourparagraph{Temporal Consistency.}
Across both datasets, the neural baselines consistently underperform compared to the best-performing \name{} configurations (see Table~\ref{tab:combined_temporal_consistency}). In addition, they are the only models that exhibit non-zero causal violations, suggesting weaker temporal consistency in their predictions. In contrast, the \name{} models substantially reduce both causal and regression violations, indicating improved temporal consistency under the proposed modeling constraints.
Figures~\ref{fig:ait_temporal_plot} and~\ref{fig:darpa_temporal_plot} further illustrate these effects. Temporal logic enforces strict ordering constraints that eliminate causal violations across all settings. The inclusion of attack-phase rules further reduces regression violations, resulting in more temporally consistent predictions overall.
Since Kitsune is an anomaly detector that only distinguishes between benign and anomalous traffic, temporal consistency metrics are not applicable and are therefore omitted.

\begin{takeaways}
\textbf{Temporal and attack-phase knowledge.}
Incorporating temporal and attack-phase knowledge improves attack-step detection performance while substantially reducing temporal inconsistencies. The strongest results are consistently obtained when both knowledge sources are combined.

\medskip

\textbf{End-to-end learning.}
Joint optimization is most beneficial when temporal constraints are complemented by informative attack-phase rules that closely align with the observed attack behavior. Otherwise, pretraining may provide more stable performance.
\end{takeaways}

\begin{figure*}[!t]
    \centering
    \captionsetup[subfigure]{font=footnotesize, skip=2pt}
    \setlength{\abovecaptionskip}{3pt}
    \setlength{\belowcaptionskip}{-4pt}

    \begin{subfigure}{0.32\linewidth}
        \centering
        \includegraphics[width=\linewidth]{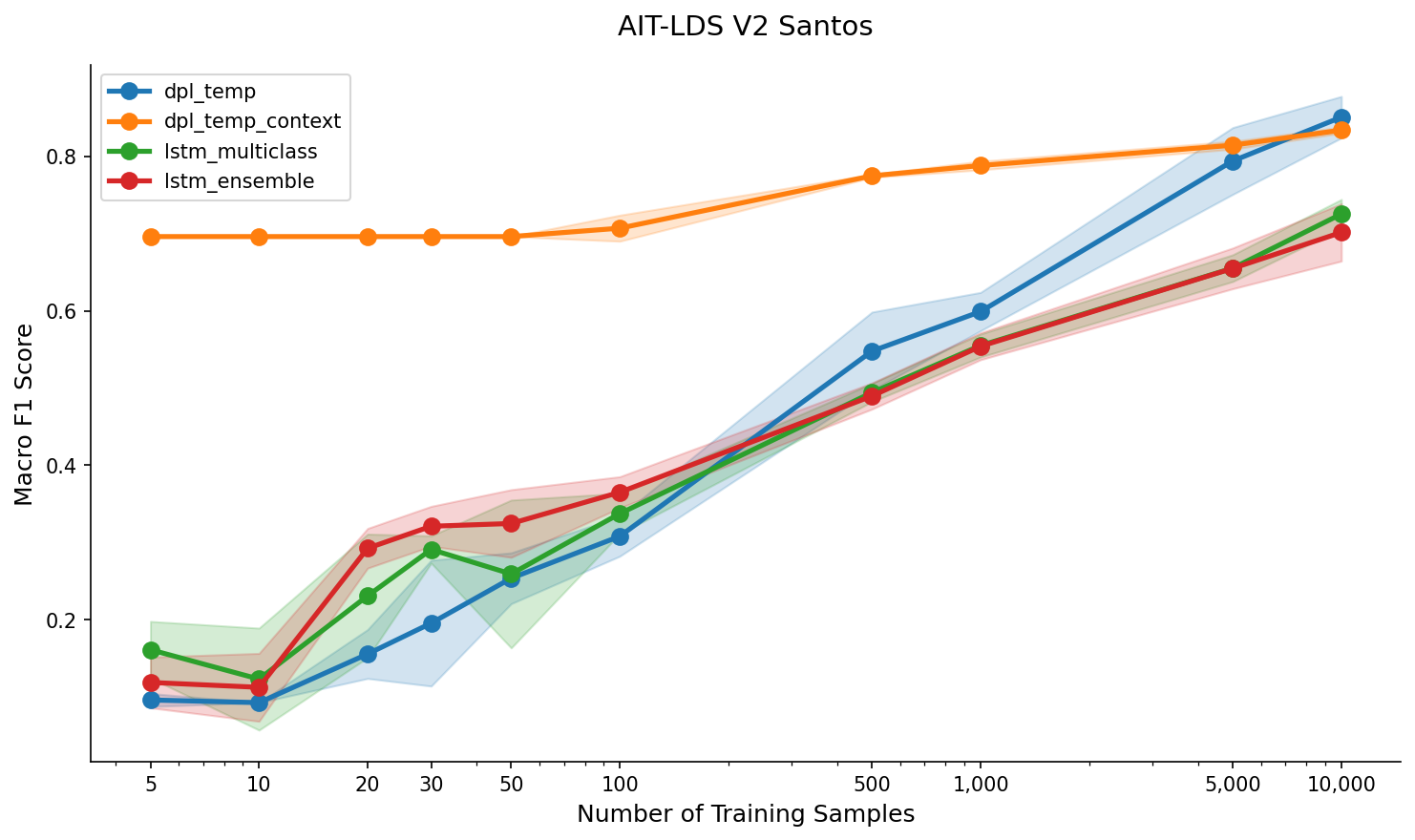}
        \caption{AIT: Macro F1}
        \label{fig:ait_macrof1}
    \end{subfigure}
    \hfill
    \begin{subfigure}{0.32\linewidth}
        \centering
        \includegraphics[width=\linewidth]{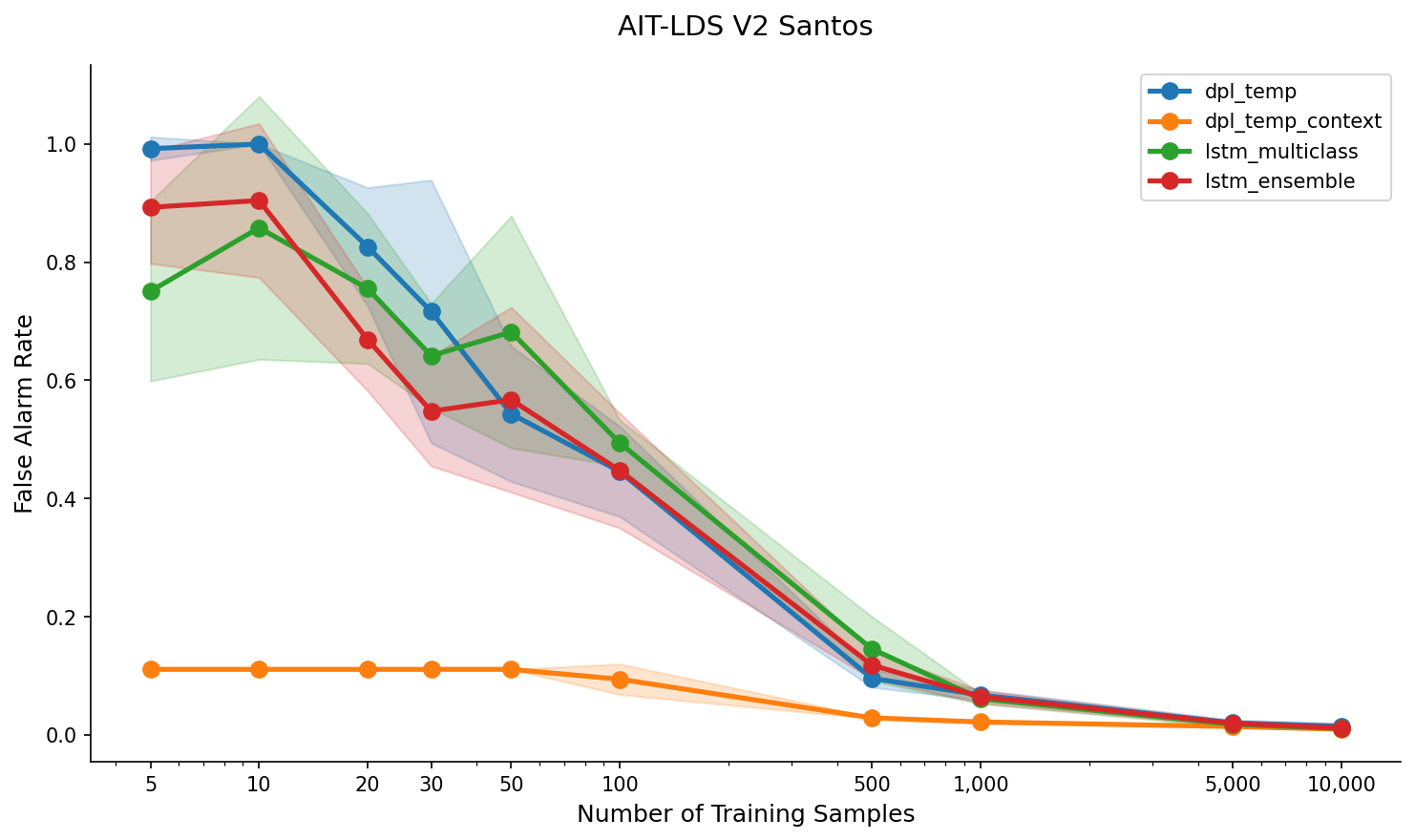}
        \caption{AIT: False alarm rate}
        \label{fig:ait_fpr}
    \end{subfigure}
    \hfill
    \begin{subfigure}{0.32\linewidth}
        \centering
        \includegraphics[width=\linewidth]{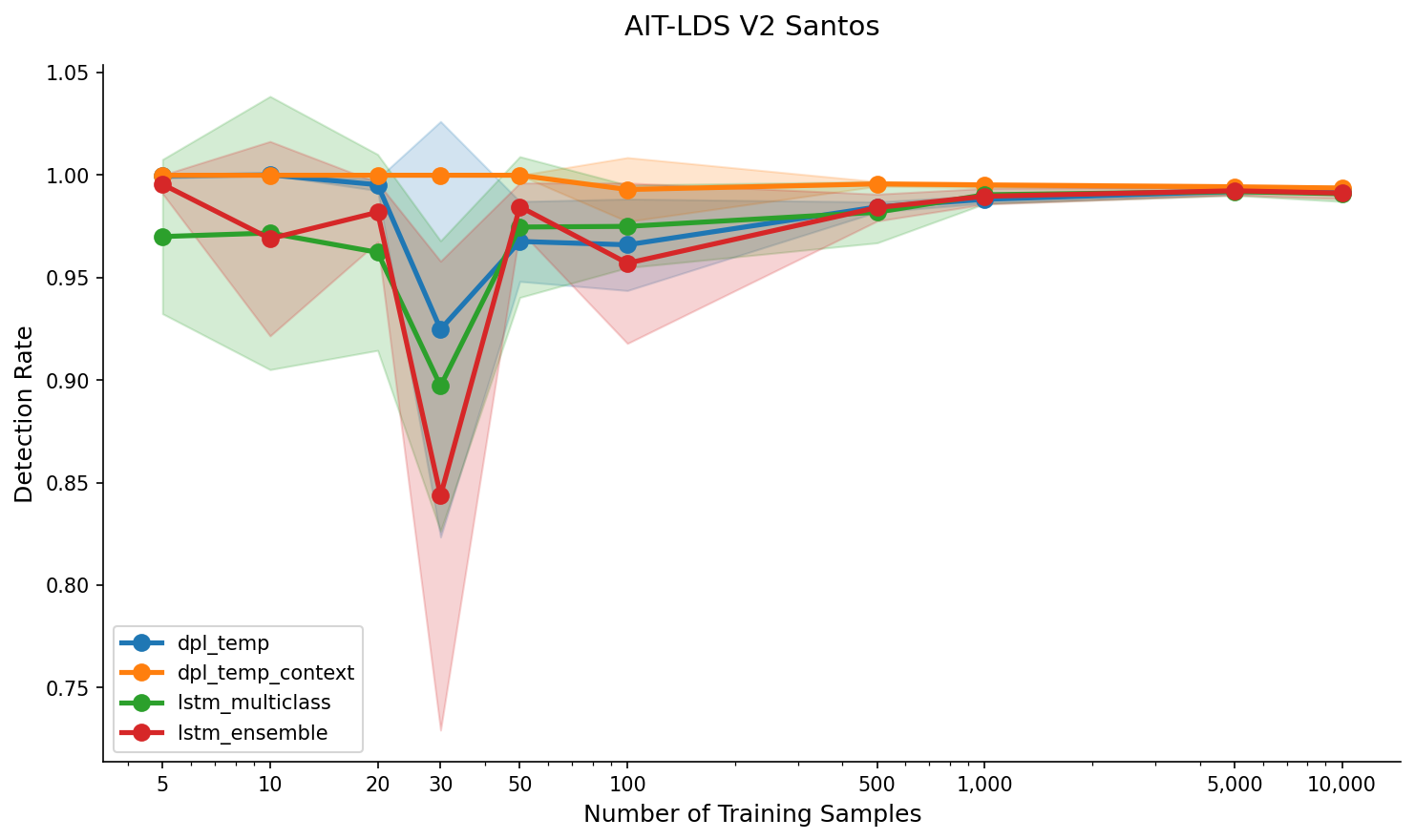}
        \caption{AIT: Detection rate}
        \label{fig:ait_dr}
    \end{subfigure}

    \vspace{0.3em}

    \begin{subfigure}{0.32\linewidth}
        \centering
        \includegraphics[width=\linewidth]{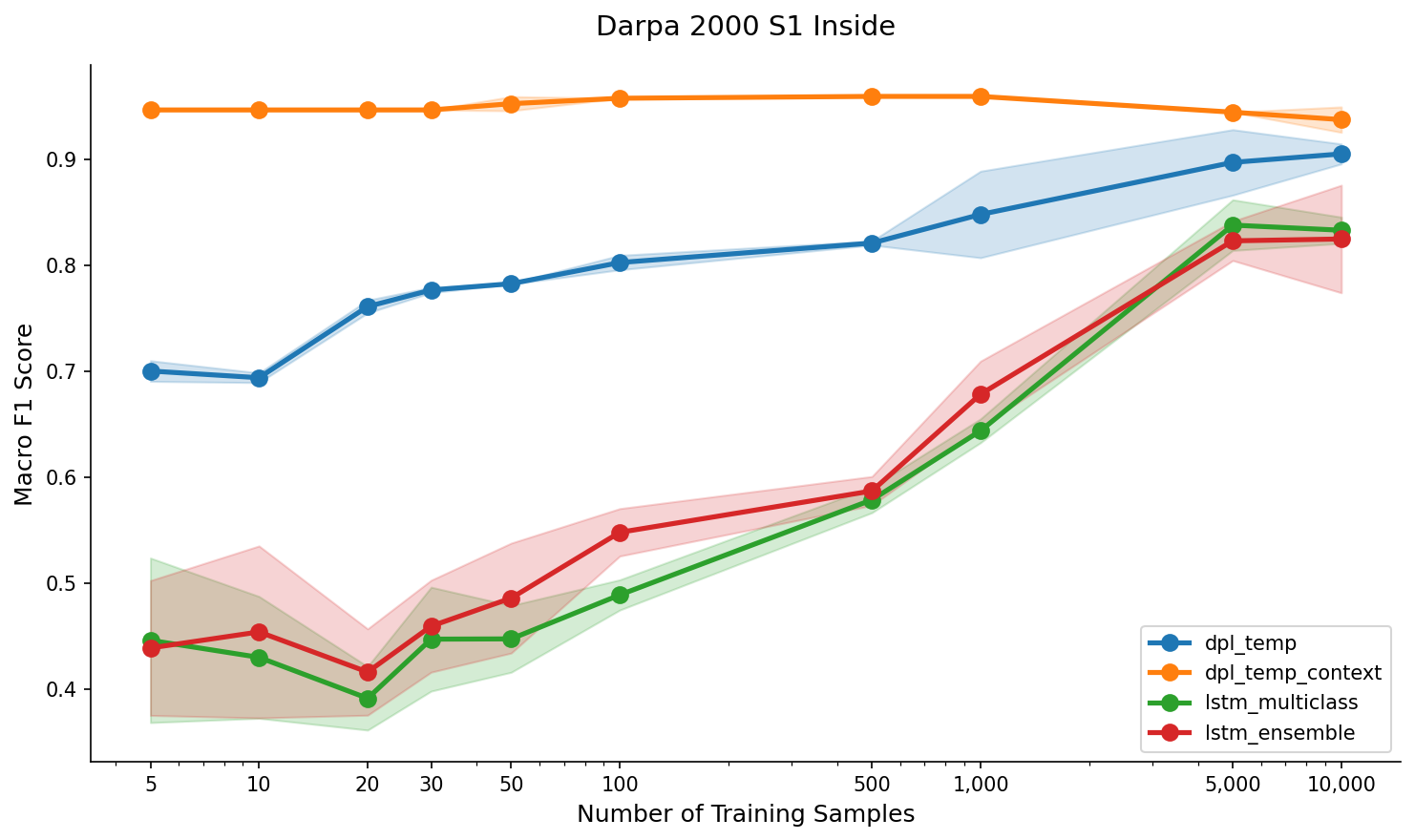}
        \caption{DARPA: Macro F1}
        \label{fig:darpa_macrof1}
    \end{subfigure}
    \hfill
    \begin{subfigure}{0.32\linewidth}
        \centering
        \includegraphics[width=\linewidth]{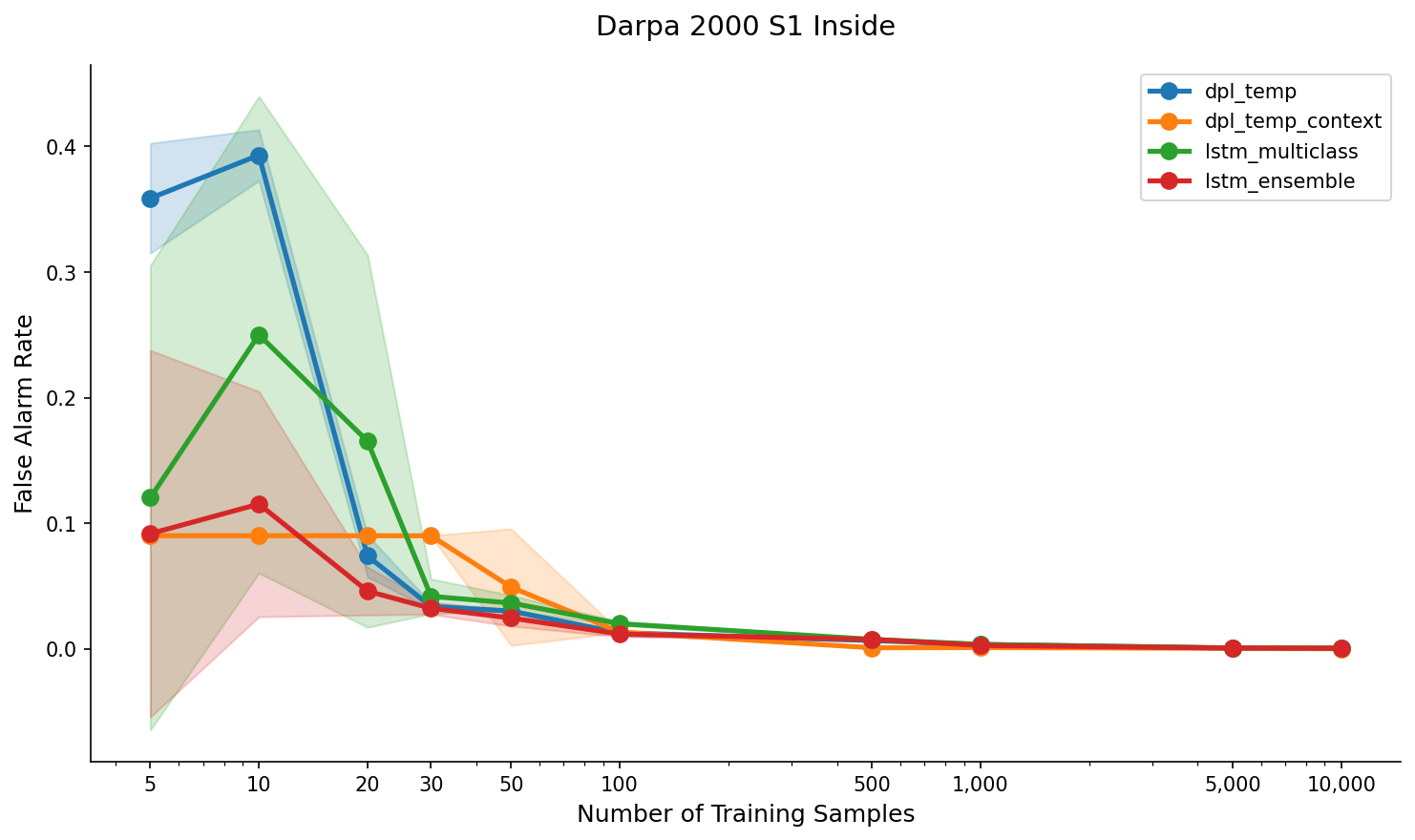}
        \caption{DARPA: False alarm rate}
        \label{fig:darpa_fpr}
    \end{subfigure}
    \hfill
    \begin{subfigure}{0.32\linewidth}
        \centering
        \includegraphics[width=\linewidth]{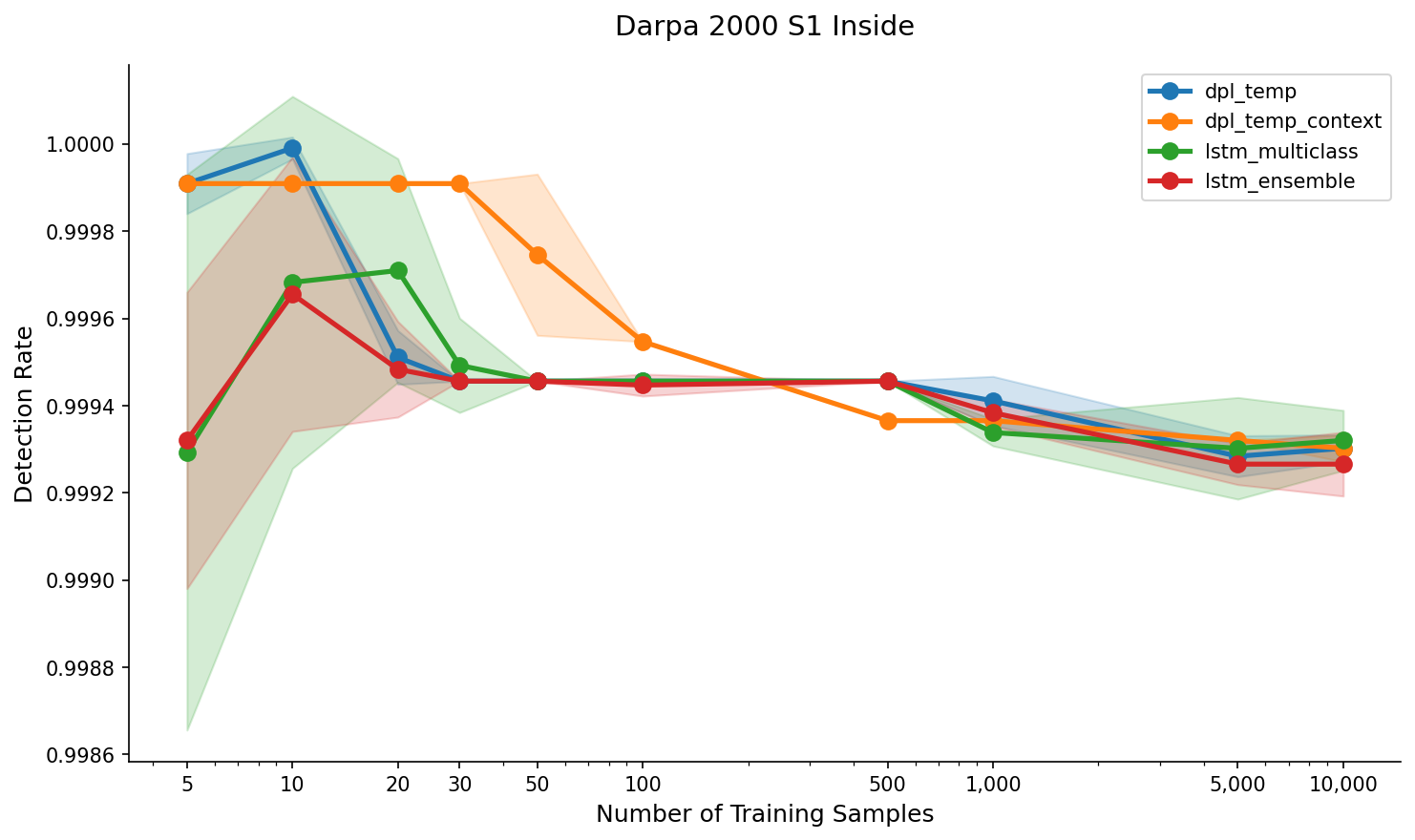}
        \caption{DARPA: Detection rate}
        \label{fig:darpa_dr}
    \end{subfigure}

    \caption{
    Effect of training set size on model performance across AIT-LDS V2 and DARPA 2000. Results show how macro F1, FAR, and DR vary as the number of training samples increases for different model configurations.
    }
    \label{fig:train_size_comparison}
\end{figure*}

\ourparagraph{Training Data Scarsity.}
We here evaluate how the performance of \name{} models and neural baselines varies under different levels of training data availability. To ensure balanced learning conditions, we construct training subsets with an equal number of benign and attack samples per attack step. For each dataset size, results are averaged over five independent runs per model.
\Cref{fig:ait_macrof1,fig:ait_fpr,fig:ait_dr} shows results for AIT-LDS V2 and \Cref{fig:darpa_macrof1,fig:darpa_fpr,fig:darpa_dr} for DARPA 2000. Across both datasets, the \name{} model with temporal and attack-phase knowledge consistently achieves strong performance in low-data regimes. In contrast, temporal knowledge alone exhibits dataset-dependent behavior. On DARPA it yields strong initial performance, while on AIT-LDS V2 it leads to slower early convergence. As the number of training samples increases, the neural baselines gradually approach the performance of the \name{} model with temporal and attack-phase knowledge. This convergence occurs at approximately 10,000 samples for AIT-LDS V2 and 5,000 samples for DARPA 2000. However, even in higher data regimes, models incorporating symbolic knowledge maintain a consistent advantage in macro F1. The error bars indicate that the \name{} model with temporal and attack-phase knowledge exhibits the most stable learning behavior across runs. In contrast, neural baselines show higher variance across all data regimes. On AIT-LDS V2, minor fluctuations are observed for the \name{} model with temporal and attack-phase knowledge at around 100 samples. A similar effect appears on DARPA at approximately 50 samples. The temporal-only \name{} model shows increased variability across most sample sizes on AIT-LDS V2, while on DARPA higher variance in macro F1 emerges beyond approximately 1,000 samples.

\begin{takeaways}
\textbf{Data efficiency.}
Temporal and attack-phase symbolic knowledge improves data efficiency, enabling strong performance in low-data regimes and reducing the number of training samples required compared to purely neural baselines.
\end{takeaways}


\ourparagraph{Generalizability.}
We analyze whether \name{} generalizes beyond the MSA scenarios it is trained on. In particular, we investigate whether incorporating symbolic knowledge enables the model to learn representations that transfer across attack variants or network vantage points, rather than overfitting to dataset-specific patterns. We focus on the \name{} model with both temporal and attack-phase knowledge---i.e., \name{} (T+R).

We consider two cross-domain evaluation settings. For AIT-LDS V2, the model is trained on one MSA scenario (Santos) and evaluated on a different scenario (Fox), allowing us to assess generalization across variations of the same MSA setting. For DARPA 2000, the model is trained on the Inside dataset and evaluated on the DMZ dataset, enabling evaluation of generalization across different network vantage points.
In both settings, models are trained on the full source scenario (Santos or Inside) and evaluated on the complete target scenario (Fox or DMZ). We compare both end-to-end training and a pretrained configuration for the neural component, and benchmark against a multiclass neural baseline.

Due to computational constraints, a window size of 10 was used for AIT-LDS V2 in this experiment instead of the default value of 100. This introduces a reduced temporal context compared to previous experiments. As a result, absolute performance levels should be interpreted with caution, and comparisons should primarily focus on relative differences between models within this setting.

\begin{figure*}[t]
    \centering

    \begin{subfigure}[t]{0.32\textwidth}
        \centering
        \includegraphics[width=\linewidth]
        {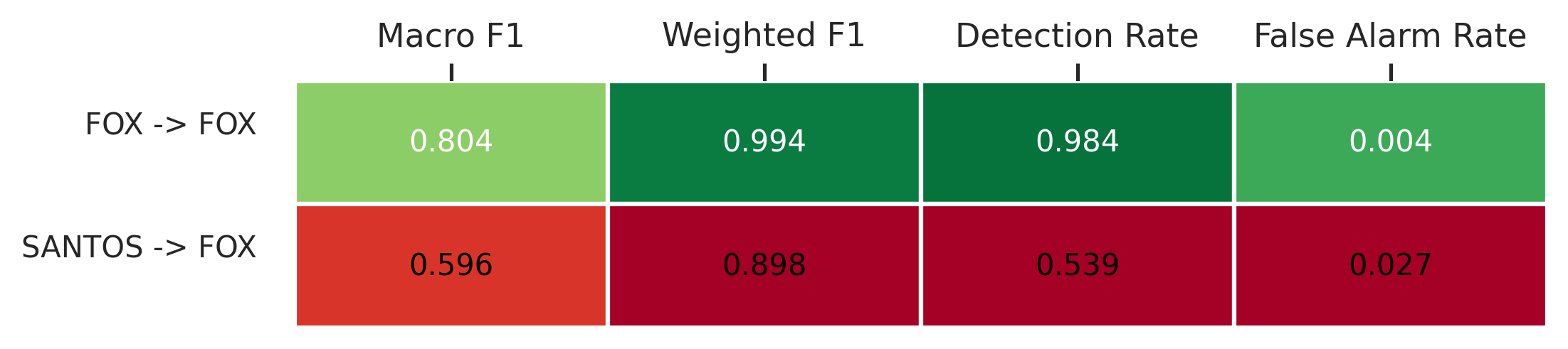}
        \caption{Multiclass LSTM}
        \label{fig:ait_neural_baseline}
    \end{subfigure}
    \hfill
    \begin{subfigure}[t]{0.32\textwidth}
        \centering
        \includegraphics[width=\linewidth]
        {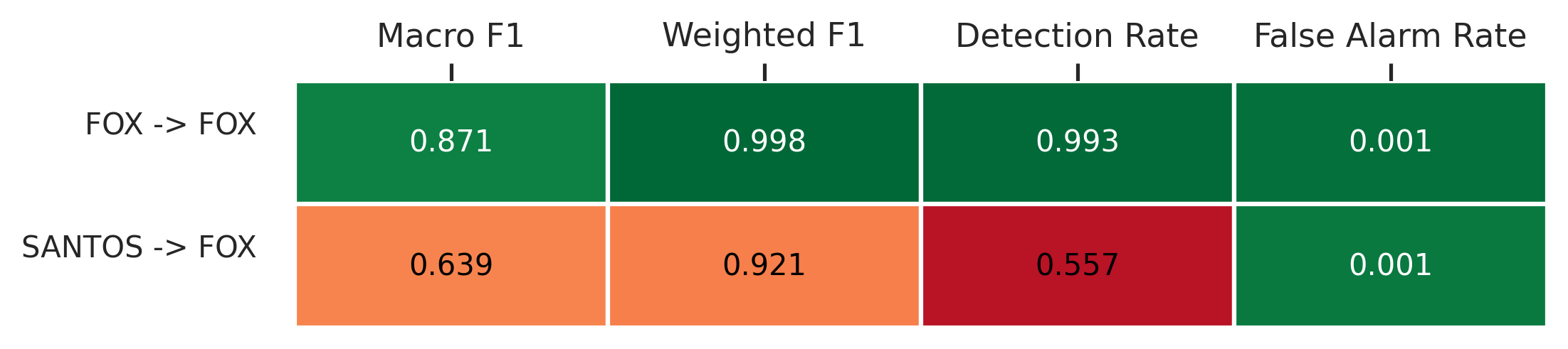}
        \caption{\name{} Pretrained}
        \label{fig:ait_dpl_pretrained}
    \end{subfigure}
    \hfill
    \begin{subfigure}[t]{0.32\textwidth}
        \centering
        \includegraphics[width=\linewidth]
        {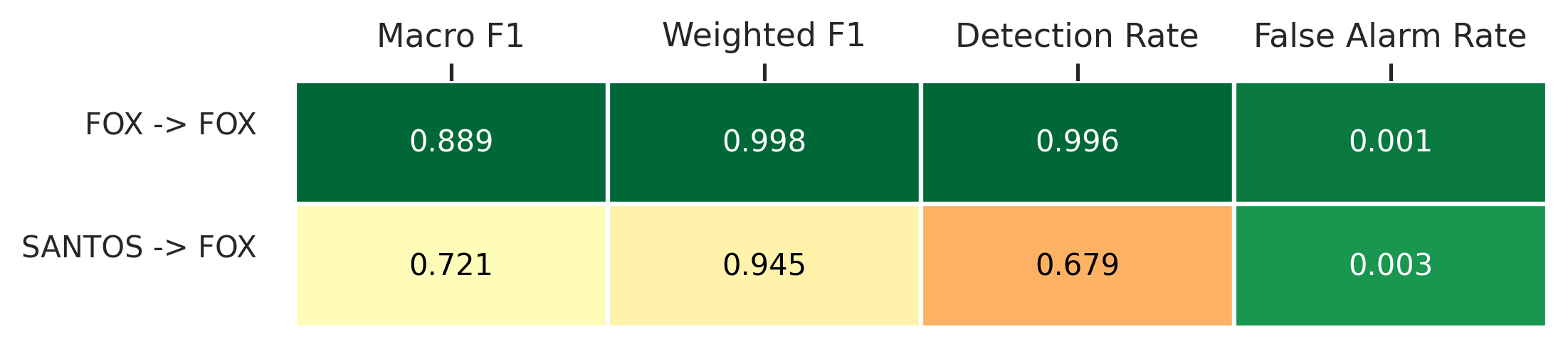}
        \caption{\name{} End-to-End}
        \label{fig:ait_dpl_endtoend}
    \end{subfigure}

    \caption{Confusion matrices illustrating cross-scenario generalization on AIT-LDSv2. Each matrix compares performance when the model is trained and tested on the same scenario against performance when the model is evaluated on a different scenario.}
    \label{fig:ait_generalizability_comparison}
\end{figure*}

Figure~\ref{fig:ait_generalizability_comparison} presents the results of generalization across scenarios on AIT-LDS V2. The neural baseline exhibits limited generalization across attack scenarios, whereas incorporating temporal and attack-phase knowledge leads to substantial improvements across all evaluation metrics. This indicates that symbolic constraints introduce useful inductive biases that transfer across attack variants. The end-to-end \name{} model achieves the strongest overall performance, suggesting that learned representations benefit from direct interaction with symbolic constraints during training.
A phase-level breakdown is provided in Table~\ref{tab:ait_generalizability}. All models struggle to detect attack phase 4 when trained on Santos and evaluated on Fox, indicating that this phase exhibits substantial variation across scenarios. Nevertheless, the \name{} models consistently outperform both the neural and symbolic baselines across most attack phases.



    
    
    
\begin{figure*}[t]
    \centering

    \begin{subfigure}[t]{0.32\textwidth}
        \centering
        \includegraphics[width=\linewidth]
        {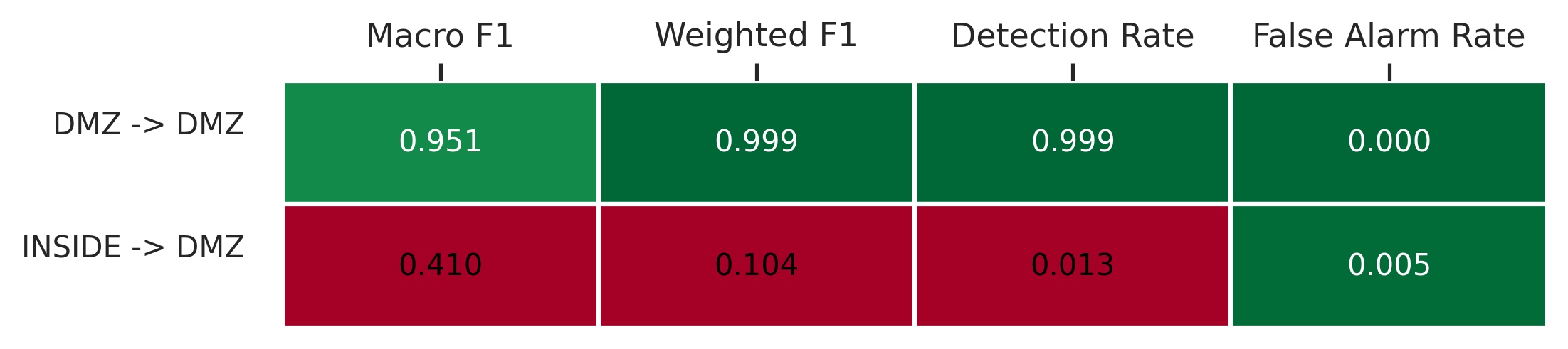}
        \caption{Multiclass LSTM}
        \label{fig:darpa_neural_baseline}
    \end{subfigure}
    \hfill
    \begin{subfigure}[t]{0.32\textwidth}
        \centering
        \includegraphics[width=\linewidth]
        {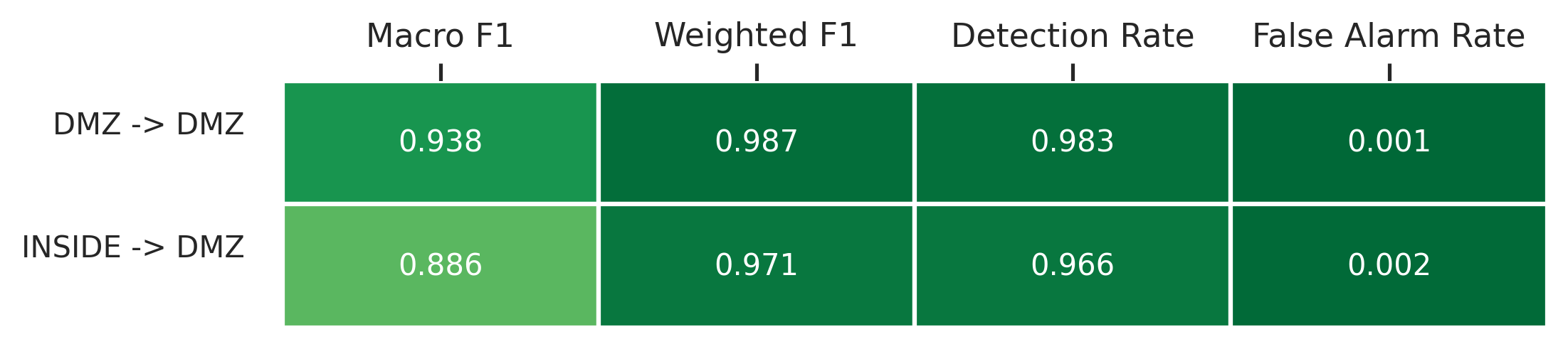}
        \caption{\name{} Pretrained}
        \label{fig:darpa_dpl_pretrained}
    \end{subfigure}
    \hfill
    \begin{subfigure}[t]{0.32\textwidth}
        \centering
        \includegraphics[width=\linewidth]
        {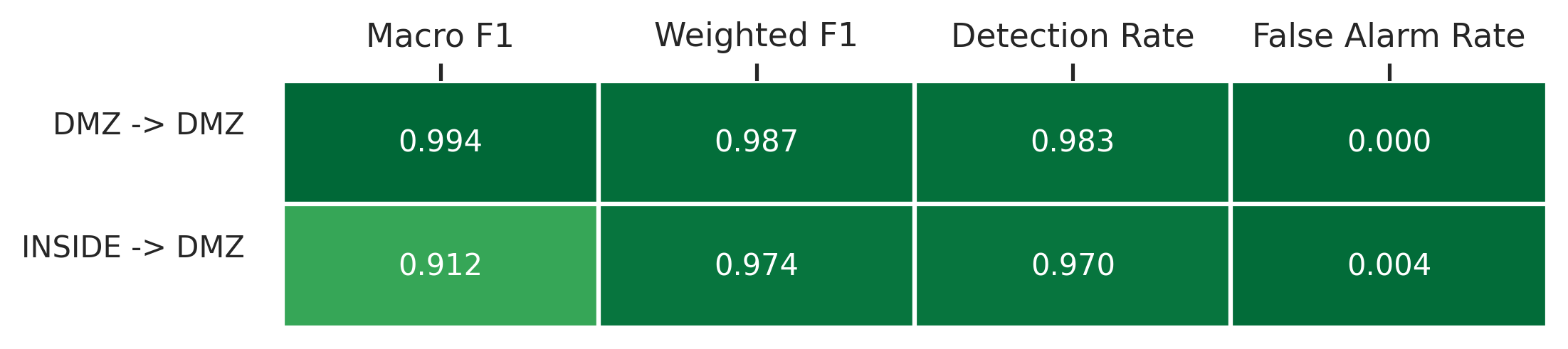}
        \caption{\name{} End-to-End}
        \label{fig:darpa_dpl_endtoend}
    \end{subfigure}

    \caption{Confusion matrices illustrating cross-sensor generalization on DARPA 2000. Each matrix compares performance when the model is trained and tested on the same sensor against performance when it is evaluated on traffic from the other sensor.}
    \label{fig:darpa_generalizability_comparison}
\end{figure*}
\begin{table}[t]
\centering
\fontsize{8}{9.5}\selectfont
\caption{Per-phase generalizability performance on AIT-LDS V2. The models are trained on Santos and tested on Fox. The best performing score per metric is in bold.}
\label{tab:ait_generalizability}
\begin{tabular}{lccccccc}
\toprule
Model & Macro F1 & Weighted F1 & F1\textsubscript{Benign} & F1\textsubscript{P1} & F1\textsubscript{P2} & F1\textsubscript{P3} & F1\textsubscript{P4} \\
\midrule
\name{}\textsubscript{E2E}        & \textbf{0.721} & 0.945 & \textbf{0.997} & 0.977 & 0.752 & \textbf{0.877} & 0.026 \\
\name{}\textsubscript{Pretrained} & 0.639 & 0.921 & 0.962 & \textbf{0.981} & 0.620 & 0.585 & \textbf{0.046} \\
LSTM\textsubscript{Multiclass}    & 0.596 & 0.898 & \textbf{0.997} & 0.965 & 0.534 & 0.511 & 0.024 \\
Logic\textsubscript{Baseline}     & 0.561 & \textbf{0.953} & 0.982 & 0.000 & \textbf{0.999} & 0.824 & 0.000 \\
\bottomrule
\end{tabular}
\end{table}

\begin{table}[t]
\centering
\fontsize{8}{9.5}\selectfont
\caption{Per-phase generalizability performance on DARPA 2000. The models are trained on INSIDE and tested on DMZ. The best performing score per metric is in bold.}
\label{tab:darpa_generalizability}
\begin{tabular}{lcccccccc}
\toprule
Model & Macro F1 & Weighted F1 & F1\textsubscript{Benign} & F1\textsubscript{P1} & F1\textsubscript{P2} & F1\textsubscript{P3} & F1\textsubscript{P4} & F1\textsubscript{P5} \\
\midrule
Logic\textsubscript{Baseline}     & \textbf{0.931} & 0.871 & 0.686 & \textbf{0.998} & \textbf{1.000} & \textbf{1.000} & \textbf{0.974} & 0.926 \\
\name{}\textsubscript{E2E}        & 0.912 & \textbf{0.974} & \textbf{0.951} & 0.586 & \textbf{1.000} & \textbf{1.000} & 0.947 & 0.990 \\
\name{}\textsubscript{Pretrained} & 0.886 & 0.971 & 0.947 & 0.431 & \textbf{1.000} & 0.974 & 0.973 & \textbf{0.991} \\
LSTM\textsubscript{Multiclass}    & 0.410 & 0.104 & 0.380 & 0.468 & 0.367 & 0.904 & 0.333 & 0.007 \\
\bottomrule
\end{tabular}
\end{table}
Figure~\ref{fig:darpa_generalizability_comparison} presents the results of the cross-vantage point experiment. Similar to the cross-scenario setting, the neural baseline shows limited generalization capability, whereas the \name{} models achieve substantially higher macro F1 and weighted F1-scores.
The highest macro F1-score is obtained by the symbolic baseline. However, this comes at the cost of a high false alarm rate, as reflected in the poor F1-score for the benign class (see Table~\ref{tab:darpa_generalizability}). In contrast, the \name{} models achieve a more balanced performance profile, while maintaining competitive overall performance.
Lastly, the neural baseline struggles across multiple phases, indicating less stable performance across attack stages.

\begin{takeaways}
\textbf{Generalization.}
Symbolic temporal and attack-phase knowledge improves generalization across both attack scenarios and network vantage points. The strongest results are consistently obtained when symbolic constraints are integrated jointly with neural learning, indicating that interaction during training is beneficial for transferability across attack variants.
\end{takeaways}

\subsection{Missed Attack Steps}
\label{ssec:missed_attack_steps}

All previous experiments assume perfect knowledge of previously observed attack phases. In practice, however, intrusion detection systems may fail to detect certain attack steps, resulting in incomplete attack histories. 
Therefore, we here investigate whether the temporal constraints can be revised to tolerate such missing observations. More specifically, we evaluate whether the model can still infer $phase_i$ even when the preceding phase $phase_{i-1}$ has not been observed. 
We relax the temporal constraint for $phase_2$ over the AIT-LDS V2 dataset, allowing the model to predict the second attack phase even when the first phase has not been observed.
%

\begin{figure}[ht]
\centering
\begin{subfigure}{0.48\linewidth}
    \centering
    \includegraphics[width=0.65\linewidth]{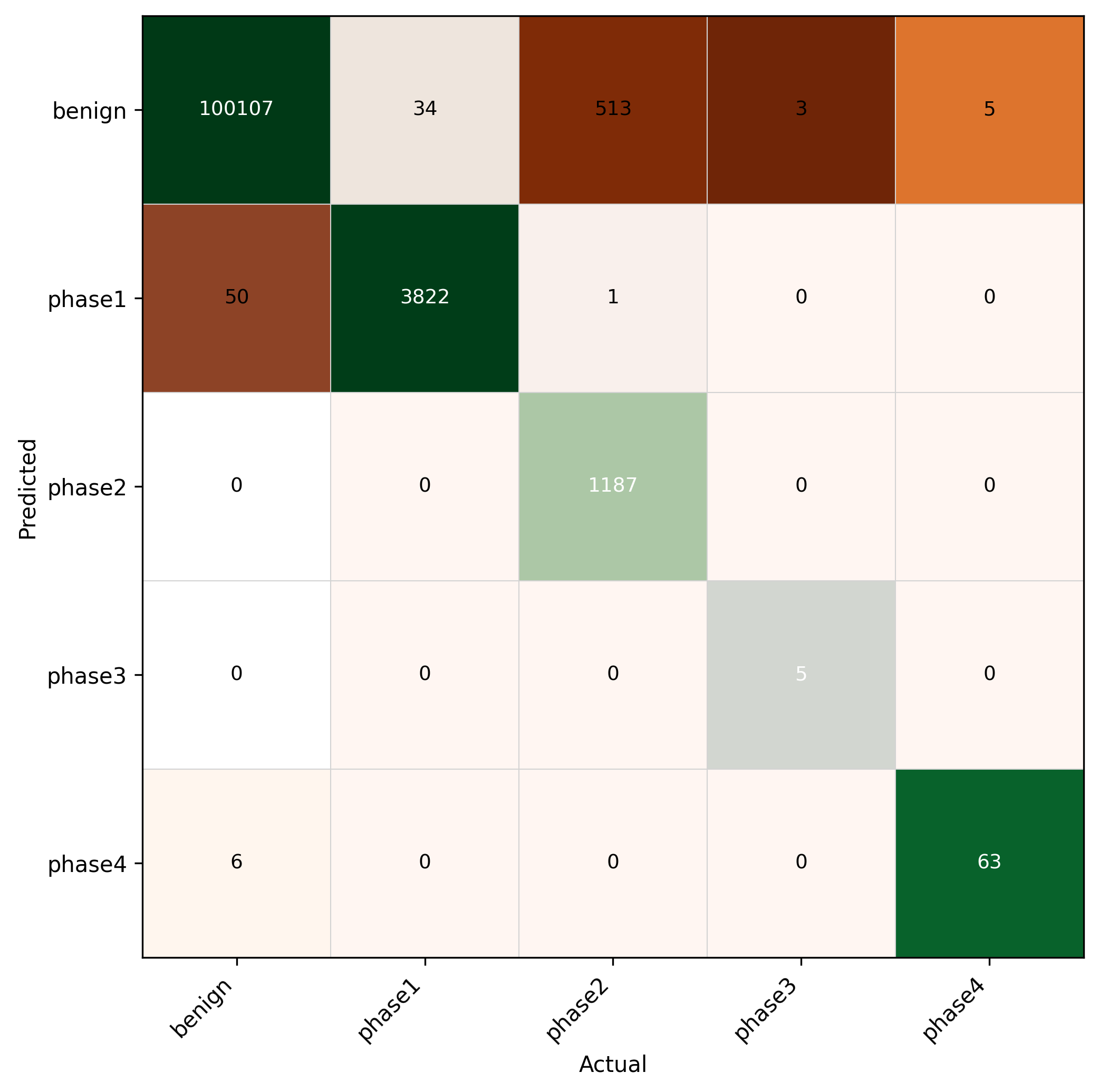}
    \caption{Detection performance of \name{} using the original logic.}
    \label{fig:ait_original_corrupted}
\end{subfigure}
\hfill
\begin{subfigure}{0.48\linewidth}
    \centering
    \includegraphics[width=0.65\linewidth]{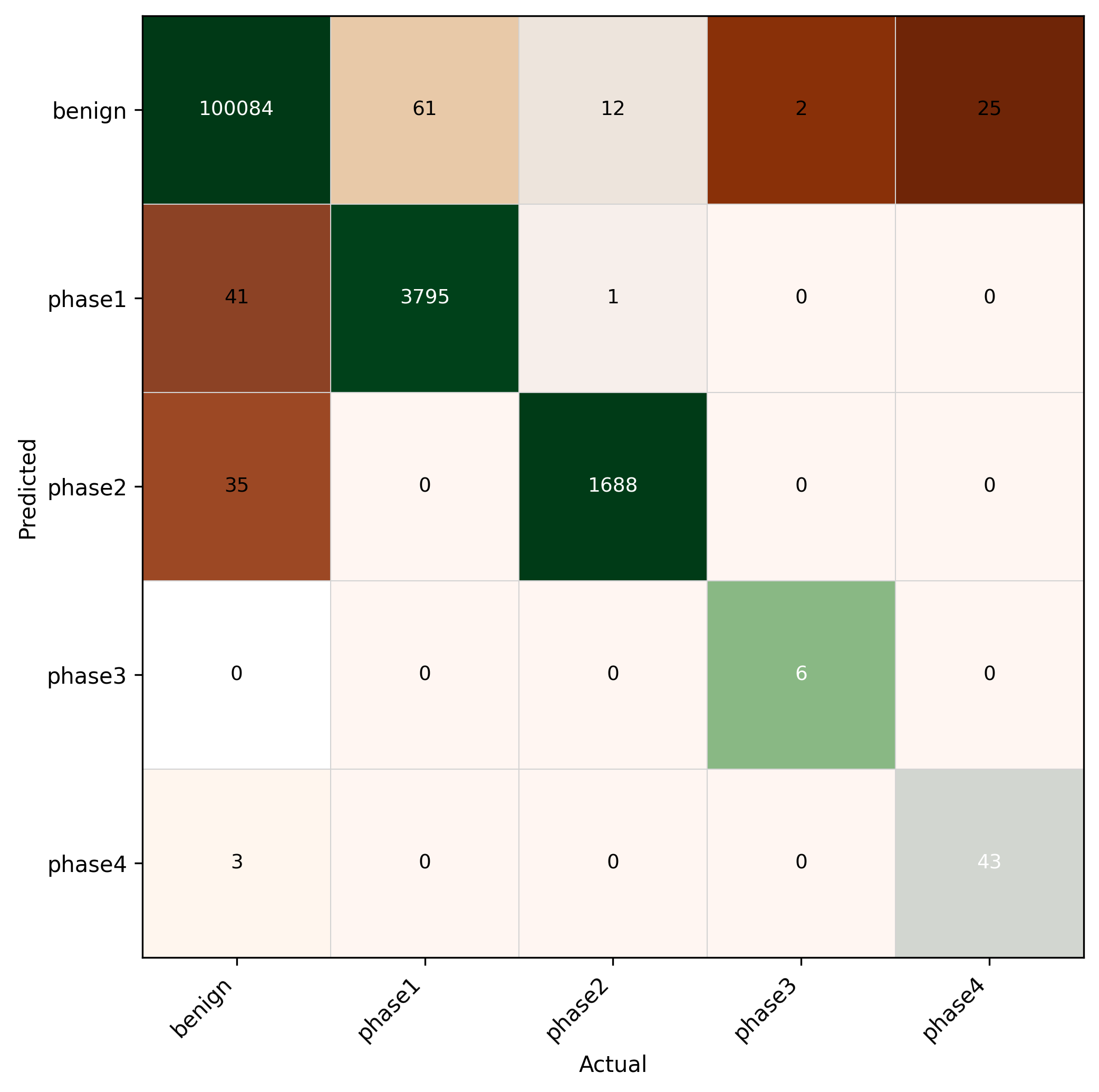}
    \caption{Detection performance of \name{} using the revised temporal logic.}
    \label{fig:ait_revised_corrupted}
\end{subfigure}

\caption{
Confusion matrices for the \name{}$_{E2E}$ model on the currepted AIT-LDS V2 dataset. Relaxing the temporal constraints used in \name{}'s symbolic knowledge defines a flexible detector that can identify attacks even when some phases observations are missed.
}
\label{fig:ait_missing_phases_comparison}
\end{figure}

To simulate missed detections, we introduce noise by randomly resetting all attack-phase flags to 0 for 30\% of the true $phase_2$ samples in both the training and test sets. This removes the temporal evidence that would normally permit the prediction of $phase_2$.
\Cref{tab:ait_missing_phases} summarizes the overall and per-phase performance. Compared to the original setting, the revised logic evaluated on the corrupted dataset exhibits a modest reduction in overall performance. Nevertheless, it substantially outperforms the original logic when applied to the same corrupted data, improving both the macro F1-score and the F1-score associated with $phase_2$.
The confusion matrices provide further insight into this behavior. As depicted in~\Cref{fig:ait_revised_corrupted}, the revised logic misses only 13 $phase_2$ flows despite the corrupted attack phase flags. In contrast, the original logic fails to recover these missing phases, resulting in 514 misclassified $phase_2$ flows (\Cref{fig:ait_original_corrupted}). 

\begin{table}[t]
\centering
\fontsize{8}{9.5}\selectfont
\caption{Performance under missing attack phases in AIT-LDS V2. The best-performing value for each metric is highlighted in bold.}
\label{tab:ait_missing_phases}
\begin{tabular}{lccccccc}
\toprule
Configuration & Macro F1 & Weighted F1 & F1\textsubscript{Benign} & F1\textsubscript{P1} & F1\textsubscript{P2} & F1\textsubscript{P3} & F1\textsubscript{P4} \\
\midrule
Logic\textsubscript{Orig}, Data\textsubscript{Orig} & \textbf{0.931} & \textbf{0.999} & \textbf{0.999} & 0.988 & \textbf{1.000} & 0.769 & 0.900 \\
Logic\textsubscript{Orig}, Data\textsubscript{Corr} & 0.899 & 0.994 & 0.997 & \textbf{0.989} & 0.822 & 0.769 & \textbf{0.920} \\
Logic\textsubscript{Rev}, Data\textsubscript{Corr}  & 0.917 & 0.998 & \textbf{0.999} & 0.987 & 0.986 & \textbf{0.857} & 0.754 \\
\bottomrule
\end{tabular}
\end{table}

\begin{takeaways}
\textbf{Robustness to missing observations.}
Relaxing temporal constraints improves robustness to missing attack observations, while preserving overall detection performance.
\end{takeaways}

\subsection{Performance Overhead}
\label{ssec:performance_overhead}

To evaluate the practical feasibility of the proposed system, we measure average inference latency per prediction, throughput, and peak memory usage. Inference time corresponds to the full execution of a \name{} query, including both logical grounding and neural forward passes when required. System metrics were collected using \texttt{psutil}  \cite{Rodola_psutil} during training and inference. 

\ourparagraph{AIT-LDS V2.}
The results in Table~\ref{tab:ait_runtime_main} show that mean inference latency ranges from 0.14 ms to 2.48 ms across models, corresponding to throughput values between approximately 400 and 7,000 predictions per second. The symbolic baselines achieve the lowest latency, followed by the neuro-symbolic models with temporal and attack-phase knowledge. These \name{} models achieve substantially lower average latency compared to their temporal-only counterparts.
Peak memory usage remains relatively stable across all configurations, ranging from approximately 25.9 GB to 30.2 GB. The small variation across models further suggests that memory consumption is largely driven by the shared execution infrastructure rather than the specific reasoning configuration.

\ourparagraph{DARPA 2000.}
As shown in Table~\ref{tab:darpa_runtime_main}, mean inference latency ranges from 2.19 ms to 5.60 ms. As in the AIT-LDS V2 setting, the symbolic baselines are the fastest, while neuro-symbolic models incur additional computational overhead due to neural execution. Overall, DARPA 2000 exhibits higher inference latency and a smaller performance gap between models compared to AIT-LDS V2.
Peak memory usage is again stable across models, ranging from approximately 12.6 GB to 15.3 GB, with only minor differences between configurations. This suggests that memory consumption scales primarily with dataset size rather than the inclusion of symbolic reasoning components.

\begin{table}[t]
\centering
\fontsize{8}{9.5}\selectfont
\caption{Runtime efficiency on the AIT-LDS V2 test set. Latency is averaged per prediction across several runs, and throughput is computed as $1/\text{latency}$. The best-performing value for each metric is highlighted in bold.}
\label{tab:ait_runtime_main}
\begin{tabular}{lccc}
\toprule
Model & \shortstack{Latency\\(ms)} & \shortstack{Throughput\\(pred/s)} & \shortstack{Peak memory\\(GB)} \\
\midrule
\name{}\textsubscript{Pretrained} T   & $2.48 \pm 40.28$ & 403  & $\mathbf{25.90 \pm 0.57}$ \\
\name{}\textsubscript{E2E} T          & $2.41 \pm 36.93$ & 415  & $29.83 \pm 0.56$ \\
\name{}\textsubscript{Pretrained} T+R & $0.46 \pm 3.32$  & 2155 & $30.19 \pm 0.01$ \\
\name{}\textsubscript{E2E} T+R        & $0.46 \pm 1.09$  & 2152 & $30.19 \pm 0.00$ \\
Logic\textsubscript{Baseline} T+R     & $0.23 \pm 15.03$ & 4354 & $30.13 \pm 0.01$ \\
Logic\textsubscript{Baseline} T       & $\mathbf{0.14 \pm 0.35}$ & \textbf{6985} & $29.95 \pm 0.03$ \\
\bottomrule
\end{tabular}
\end{table}

\begin{table}[t]
\centering
\fontsize{8}{9.5}\selectfont
\caption{Runtime efficiency on the DARPA 2000 test set. Latency is averaged per prediction across several runs, and throughput is computed as $1/\text{latency}$. The best-performing value for each metric is highlighted in bold.}
\label{tab:darpa_runtime_main}
\begin{tabular}{lccc}
\toprule
Model & \shortstack{Latency\\(ms)} & \shortstack{Throughput\\(pred/s)} & \shortstack{Peak memory\\(GB)} \\
\midrule
\name{}\textsubscript{Pretrained} T   & $5.28 \pm 36.53$ & 189 & $15.32 \pm 0.02$ \\
\name{}\textsubscript{E2E} T          & $5.34 \pm 34.69$ & 187 & $15.31 \pm 0.00$ \\
\name{}\textsubscript{Pretrained} T+R & $5.60 \pm 38.83$ & 178 & $15.19 \pm 0.00$ \\
\name{}\textsubscript{E2E} T+R        & $5.29 \pm 40.86$ & 189 & $15.20 \pm 0.01$ \\
Logic\textsubscript{Baseline} T+R     & $3.06 \pm 31.49$ & 327 & $\mathbf{12.65 \pm 0.64}$ \\
Logic\textsubscript{Baseline} T       & $\mathbf{2.19 \pm 22.00}$ & \textbf{457} & $12.99 \pm 0.01$ \\
\bottomrule
\end{tabular}
\end{table}

\begin{takeaways}
\textbf{Inference latency and memory consumption.}
The proposed neuro-symbolic approach achieves low inference latency and stable memory consumption. 
\end{takeaways}
\section{Discussions and Limitations}
\label{sec:discussions_limitations}

\ourparagraph{Complementary Neural and Symbolic Reasoning.}
The results show that neural learning and symbolic reasoning provide complementary evidence for MSA detection. Attack-phase rules assess whether flows exhibit phase-compatible characteristics, while temporal constraints enforce consistency with the observed attack progression. Symbolic reasoning is particularly effective on DARPA 2000, whose phases have well-defined network characteristics, whereas neural learning is more important on AIT-LDSv2, where attack and benign behavior overlap. Their combination generally performs best because learned patterns compensate for incomplete rules, while symbolic knowledge suppresses implausible predictions. This benefit nevertheless depends on rule quality. Restrictive rules may exclude valid variants, whereas broad rules provide limited discrimination.

\ourparagraph{Temporal Reasoning, Data Efficiency, and Attack History.}
\name{} separates local flow classification from long-range progression reasoning: the LSTM models recent traffic, while the symbolic history relates predictions across longer periods. This allows dataset-specific neural windows --\(100\) flows for AIT-LDSv2 and \(10\) for DARPA 2000 -- without limiting the duration modeled symbolically, and is particularly beneficial when labeled data are scarce. Similar distinctions between low-level evidence and higher-level attack stories appear in SLEUTH, HOLMES, and ATLAS \cite{hossain2017sleuth,milajerdi2019holmes,alsaheel2021atlas}. However, the evaluation assumes one predefined progression and uses ground-truth phases to update the history. It therefore excludes phase reordering, interleaved attacks, competing hypotheses, and error propagation from previous predictions. The results consequently quantify the value of temporal knowledge under controlled-history conditions rather than fully online attack reconstruction.

\ourparagraph{Training Strategy and Computational Cost.}
Pretrained \name{} models demonstrate that symbolic knowledge improves inference even when introduced after neural training, while the stronger cross-dataset performance of end-to-end models suggests that joint optimization better aligns learned representations with the logic program. Symbolic inference introduces modest overhead, and attack-phase rules can reduce latency by excluding incompatible flows before neural evaluation. Nevertheless, the measurements cover model inference rather than the complete operational pipeline; online flow extraction, query construction, state management, larger rule bases, concurrent attacks, and sustained traffic may introduce additional costs.

\ourparagraph{Observability, Generalization, and Evasion.}
The evaluation uses network-flow metadata from two controlled benchmarks and therefore captures only the network-observable portion of an attack. Host-local actions, including privilege escalation, credential access, persistence, and command execution, may require richer telemetry; provenance-based systems such as HOLMES and UNICORN obtain such evidence from host auditing at the cost of more invasive data collection \cite{milajerdi2019holmes,han2020unicorn}. Moreover, performance on DARPA 2000 and AIT-LDSv2 does not establish effectiveness on operational traffic with different base rates, environments, and attacker behavior \cite{sommer2010outside}. Adaptive attackers may also manipulate flow characteristics, reorder phases, insert benign activity, or contaminate the stored history. Practical evasion and mimicry attacks against learned and provenance-based detectors have been demonstrated \cite{pierazzi2020intriguing,goyal2023mimicry}. Robustness against such behavior therefore requires dedicated adversarial evaluation.

\section{Related Work}
\label{sec:related_work}

Prior MSA defenses largely follow a \emph{detect first, correlate later} paradigm where IDS first identify suspicious events, after which a separate stage links them into an attack scenario~\cite{navarro2018survey}.

\ourparagraph{Detect First, Correlate Later.}
Classical alert-correlation methods link IDS outputs through shared attributes, causal dependencies, or attack graphs.
Ning et al. connect alerts when the consequences of one action satisfy the prerequisites of another and later hypothesize steps missed by the IDS~\cite{ning2002constructing,ning2004building}.
Angelini et al. detect online progress along known attack-graph paths, while M$^2$ASK maps NIDS alerts to MITRE ATT\&CK and correlates them into multi-step scenarios~\cite{angelini2018attackGraph,meng2024m2ask}.
These systems may operate online, but progression reasoning remains downstream of alert generation and therefore cannot revise the detection of the underlying events.

Provenance-based systems similarly reason over host audit events after their collection.
For example, HOLMES correlates suspicious information flows into high-level APT scenarios~\cite{milajerdi2019holmes}, while UNICORN detects anomalous long-running provenance graphs~\cite{han2020unicorn}, and MAGIC uses self-supervised masked graph learning for multi-granularity APT detection and concept-drift adaptation~\cite{jia2024magic}.
NODLINK performs online APT detection by screening provenance events and approximating a Steiner tree that recovers a concise attack-related subgraph~\cite{li2024nodlink}.
NODLINK is closest to \name{} in targeting timely detection of evolving attacks, but differs in observability and output as it requires endpoint provenance and reconstructs a causal investigation graph, whereas \name{} uses network flows and returns progression-consistent attack phase alerts.
The approaches are therefore complementary rather than directly interchangeable: provenance provides richer host-level causality, while flow-level detection avoids endpoint instrumentation and remains applicable to encrypted network traffic.


\ourparagraph{Learning Attack Progression.}
Data-driven IDSs such as DeepLog and Kitsune learn anomalous behavior from logs or flows, but do not explicitly represent an attack's phase progression~\cite{du2017deeplog,mirsky2018kitsune}.
More recent systems aim to model longer contexts.
For example,~\cite{ghafir2018mlapt} predicts APT progression from specialized detector outputs, while StageFinder estimates ATT\&CK stages from provenance graphs~\cite{phan2026stagefinder}, and Tracegram learns temporal dependencies among ordered flows~\cite{qu2026tracegram}.
These methods demonstrate the value of temporal context, but either depend on upstream detections or learn progression statistically without enforcing explicit phase prerequisites. 

\name{} overcomes these limitations through \emph{detection through logical progression}.
Instead of learning attack progression statistically, \name{} embeds explicit attack-phase rules and temporal prerequisites directly into flow-level inference using neuro-symbolic reasoning.  
Progression knowledge is therefore explicitly encoded and constrains each detection decision with complete trustworthiness. 

\section{Conclusions}
\label{sec:conclusions}

We presented \name{}, a neuro-symbolic NIDS that embeds temporal and attack-phase knowledge directly into attack detection. 
By combining neural classification, probabilistic symbolic reasoning, and context-specific attack memory, \name{} produces predictions that are both data-driven and consistent with the modeled attack progression. 
Evaluation on DARPA 2000 and AIT-LDS V2 shows that \name{} improves attack-phase classification over neural baselines, eliminates causal violations, and improves data efficiency and cross-domain generalization. 
These results demonstrate the potential of neuro-symbolic reasoning to support more coherent and operationally useful detection of multi-step attacks.

\bibliographystyle{plain}
\bibliography{references}  





\appendix
\section{Use of Generative AI}
\label{secapp:use_genai}
Generative AI tools -- mainly GPT 5.6 -- were used to improve the English language, clarity, and readability of the manuscript. Their use was limited primarily to grammar correction, sentence restructuring, and stylistic refinement. The authors developed the research methodology, implemented the system, conducted and interpreted the experiments, and wrote the substantive content of the manuscript. All AI-assisted text was reviewed, revised, and approved by the authors, who take full responsibility for the final content.

\section{Dataset Details}
\label{secapp:dataset_details}

\subsection{DARPA 2000 Intrusion Detection Dataset} 
\label{app:darpa2000}
DARPA 2000 was produced by MIT Lincoln Laboratory and captures network traffic from a simulated Air Force network. Packet captures are available from sensors located inside the network and in the demilitarized zone (DMZ). The dataset contains two distributed denial-of-service (DDoS) scenarios, i.e., LLDOS~1.0, executed by a novice attacker, and the stealthier LLDOS~2.0.2 scenario. Because LLDOS~1.0 contains substantially more attack observations, we use its inside and DMZ traffic for training and evaluation. The LLDOS~1.0 attack consists of five phases \cite{DARPA2000}: 
\begin{enumerate} 
\item \textbf{Scanning:} ICMP Echo Requests are used to discover active hosts. 

\item \textbf{Probing:} \texttt{sadmind} ``ping'' requests identify Solaris hosts running the vulnerable service. 

\item \textbf{Exploitation:} the attacker exploits the \texttt{sadmind} vulnerability, CVE-1999-0977 \cite{CVE19990977}, to compromise selected hosts. 

\item \textbf{Installation:} the \texttt{mstream} DDoS software is installed on the compromised hosts. 

\item \textbf{DDoS:} the compromised hosts launch a distributed denial-of-service attack against an external target. 
\end{enumerate} 
\subsection{AIT Log Dataset V2} \label{app:ait_ldsv2}
AIT Log Dataset V2 (AIT-LDSv2) comprises eight synthetic datasets collected from enterprise-like testbeds at the Austrian Institute of Technology in 2022 \cite{landauer2021have,landauer2023maintainable}. Each dataset covers four to six days of benign activity and a multi-step attack executed during the observation period. The testbeds contain multiple hosts, services, and users, with network traffic and host-level logs collected throughout the environment. The attack comprises six phases \cite{landauer2021have,landauer2023maintainable}: 
\begin{enumerate} 
\item \textbf{Scanning:} host discovery, DNS brute-force, service scanning, directory enumeration, and WordPress scanning are performed using tools including Nmap, Dirb, and WPScan. 

\item \textbf{Exploitation:} CVE-2020-24186 is exploited to upload a PHP webshell, which is subsequently used for host discovery and command execution. 

\item \textbf{Password Cracking:} password hashes are attacked using John the Ripper and a common-password wordlist. 

\item \textbf{Privilege Escalation:} a compromised account and an interactive reverse shell are used to obtain root privileges. 

\item \textbf{Remote Command Execution:} the attacker executes commands, including operations requiring root privileges. 

\item \textbf{Data Exfiltration:} DNSteal is used to exfiltrate data from an intranet file share. \end{enumerate} 
Although the eight scenarios implement the same attack, they vary in scan order, duration, event frequency, and phase coverage. We use the \textit{fox} and \textit{santos} scenarios, which respectively exhibit high and low scan volumes. This contrast enables evaluation under different traffic and attack-frequency conditions. Table~\ref{tab:ait_scenarios} summarizes the selected scenarios. 

\begin{table}[t]
\centering
\fontsize{8}{9.5}\selectfont
\caption{Selected AIT-LDS V2 scenarios. Adapted from \cite{AITLDSV2}.}
\label{tab:ait_scenarios}
\begin{tabular}{lllll}
\toprule
Scenario & Simulation period & Attack period & Scan volume & Size \\
\midrule
\textit{fox}    & 2022-01-15 to 2022-01-20 & 2022-01-18 11:59--13:15 & High & 26\,GB \\
\textit{santos} & 2022-01-14 to 2022-01-18 & 2022-01-17 11:15--11:59 & Low  & 17\,GB \\
\bottomrule
\end{tabular}
\end{table}
To the best of our knowledge, this is the first evaluation of multi-step attack detection on AIT-LDSv2. 

\ourparagraph{Network-Observable Attack Sequence.} 
AIT-LDSv2 was designed primarily for log-based analysis.
Consequently, not every labeled attack phase is observable from network traffic alone. In particular, privilege escalation and remote command execution rely predominantly on host-level evidence. Because our evaluation uses only network traffic, these phases are excluded from the modeled attack sequence. Data exfiltration also differs from the expected logical progression. The dataset authors start DNSteal near the beginning of the simulation to contaminate the nominal training period and challenge anomaly-detection methods that assume attack-free training data \cite{landauer2023maintainable}. Exfiltration therefore precedes the remaining attack phases in the recorded timeline, rather than representing the attacker's final objective. We preserve the dataset's observed temporal order and evaluate the following network-visible sequence: 
\begin{inlinelist} 
\item \textbf{Data Exfiltration},
\item \textbf{Scanning},
\item \textbf{Exploitation}, and 
\item \textbf{Password Cracking}.
\end{inlinelist}

\section{Data Preparation} \label{app:data_preparation}
The data preparation pipeline comprises four stages: flow extraction, attack-phase labeling, neural input construction, and \name{} query generation. 
\subsection{Flow Extraction and Labeling} \label{app:flow_processing}
Raw PCAP traffic is converted into flow-level records using the Zeek network security monitor \cite{paxson1999bro}. Zeek produces several protocol-specific logs.
However, because \name{} assumes that packet payloads may be encrypted, we use only \texttt{conn.log}, which contains connection metadata derived from packet headers. 
\paragraph{DARPA 2000.} The dataset provides a complete traffic capture containing both benign and malicious activity, together with separate PCAP subsets for the five attack phases. Zeek is applied uniformly to all captures, and flows are matched using hashes computed over their connection attributes, excluding the unique Zeek identifier. A flow in the complete capture receives the label of its matching phase-specific flow; unmatched flows are labeled as benign. Table~\ref{tab:label_distribution_darpa} reports the resulting distributions for the inside and DMZ sensors. 
\begin{table}[t]
\centering
\fontsize{8}{9.5}\selectfont
\caption{Flow-label distribution for the DARPA 2000 LLDOS~1.0 scenario.}
\label{tab:label_distribution_darpa}
\begin{tabular}{lrrrrrr}
\toprule
Sensor & Benign & Phase\textsubscript{1} & Phase\textsubscript{2} & Phase\textsubscript{3} & Phase\textsubscript{4} & Phase\textsubscript{5} \\
\midrule
Inside & 52,258 & 20  & 22 & 34 & 19 & 73,472 \\
DMZ    & 10,609 & 767 & 25 & 79 & 19 & 33,942 \\
\bottomrule
\end{tabular}
\end{table}
\paragraph{AIT-LDSv2.} The PCAP files collected from different sensors are processed separately with Zeek and merged according to connection start time. We retain traffic within the official simulation period and remove duplicate flows reported by the same sensor with identical attributes \cite{landauer2021have,landauer2023maintainable}. AIT-LDSv2 provides labels for flows extracted using Tstat. Because Tstat and Zeek produce different flow representations, labels cannot be transferred through exact record matching. We instead align the records using their source and destination addresses, source and destination ports, and temporal proximity. The original fine-grained labels are subsequently mapped to the attack phases modeled in this work using the accompanying labeling notebook. Table~\ref{tab:label_distribution_ait} shows the resulting distributions. 
\begin{table}[t]
\centering
\fontsize{8}{9.5}\selectfont
\caption{Flow-label distribution for the selected AIT-LDS V2 scenarios.}
\label{tab:label_distribution_ait}
\begin{tabular}{lrrrrr}
\toprule
Scenario & Benign & Phase\textsubscript{1} & Phase\textsubscript{2} & Phase\textsubscript{3} & Phase\textsubscript{4} \\
\midrule
\textit{fox}    & 400,024 & 14,678 & 56,443 & 28 & 61  \\
\textit{santos} & 333,964 & 12,861 & 5,670  & 26 & 228 \\
\bottomrule
\end{tabular}
\end{table}
\subsection{Neural Input Construction} \label{app:neural_input_preparation}
Neural inputs combine connection-level metadata from \texttt{conn.log} with statistical features computed from recent network activity.
Table~\ref{tab:zeek_features} summarizes the selected Zeek attributes and their use in \name{}. The unique connection identifier \texttt{uid} is discarded, \texttt{history} is excluded to avoid a high-dimensional categorical encoding, and \texttt{tunnel\_parents} is omitted because it is absent from the evaluated datasets. 
\begin{table}[t]
\centering
\fontsize{8}{9.5}\selectfont
\caption{Zeek \texttt{conn.log} attributes used by \name{}.}
\label{tab:zeek_features}
\begin{tabular}{lll}
\toprule
Attribute & Description & Usage \\
\midrule
\texttt{ts}             & Connection timestamp           & Symbolic \\
\texttt{id.orig\_h}     & Source address                 & Symbolic \\
\texttt{id.resp\_h}     & Destination address            & Symbolic \\
\texttt{id.orig\_p}     & Source port                    & Symbolic \\
\texttt{id.resp\_p}     & Destination port               & Symbolic \\
\addlinespace
\texttt{local\_orig}    & Internal source indicator      & Symbolic + Neural \\
\texttt{local\_resp}    & Internal destination indicator & Symbolic + Neural \\
\texttt{proto}          & Transport protocol             & Symbolic + Neural \\
\addlinespace
\texttt{service}        & Identified application service & Neural \\
\texttt{duration}       & Connection duration            & Neural \\
\texttt{orig\_bytes}    & Originator payload bytes       & Neural \\
\texttt{resp\_bytes}    & Responder payload bytes        & Neural \\
\texttt{conn\_state}    & Connection state               & Neural \\
\texttt{missed\_bytes}  & Bytes missed during capture    & Neural \\
\texttt{orig\_pkts}     & Originator packet count        & Neural \\
\texttt{orig\_ip\_bytes}& Originator IP bytes            & Neural \\
\texttt{resp\_pkts}     & Responder packet count         & Neural \\
\texttt{resp\_ip\_bytes}& Responder IP bytes             & Neural \\
\texttt{ip\_proto}      & IP protocol number             & Neural \\
\addlinespace
\texttt{uid}            & Unique connection identifier   & Excluded \\
\texttt{history}        & Packet-history flags           & Excluded \\
\texttt{tunnel\_parents}& Parent tunnel identifiers      & Excluded \\
\bottomrule
\end{tabular}
\end{table}

\paragraph{Statistical Features.} For every flow, we summarize the preceding 60 seconds of traffic. Source-oriented features comprise the number of initiated connections, unique destination addresses, and unique destination ports. Destination-oriented features comprise the number of incoming connections and unique source addresses. We additionally compute the proportions of connections exhibiting SYN failures, rejections, and resets. The 60-second aggregation interval was selected experimentally and is configurable: longer intervals capture slower behaviors but may also introduce unrelated traffic. 

\paragraph{Feature Processing.} Categorical features are one-hot encoded, whereas numerical features are scaled to the interval $[0,1]$ using MinMax normalization. Table~\ref{tab:feature_type} lists the resulting neural feature groups. 
\begin{table}[t]
\centering
\fontsize{8}{9.5}\selectfont
\caption{Features provided to the neural component.}
\label{tab:feature_type}
\begin{tabularx}{\linewidth}{@{}l>{\raggedright\arraybackslash}X@{}}
\toprule
Type & Features \\
\midrule
Categorical & \texttt{local\_orig}, \texttt{local\_resp}, \texttt{proto}, \texttt{service}, \texttt{conn\_state}, and \texttt{ip\_proto}. \\
\addlinespace
Numerical & \texttt{duration}, \texttt{orig\_bytes}, \texttt{resp\_bytes}, \texttt{missed\_bytes}, \texttt{orig\_pkts}, \texttt{orig\_ip\_bytes}, \texttt{resp\_pkts}, \texttt{resp\_ip\_bytes}, \texttt{connections\_per\_src}, \texttt{unique\_targets}, \texttt{unique\_dports}, \texttt{connections\_per\_dst}, \texttt{unique\_sources\_per\_dst}, \texttt{syn\_failure\_ratio}, \texttt{reject\_ratio}, and \texttt{reset\_ratio}. \\
\bottomrule
\end{tabularx}
\end{table}

\paragraph{Temporal Windows.} The transformed flows are ordered chronologically and grouped into fixed-length sequences. Each sequence contains the current flow and its preceding flows and is used to predict the phase associated with the current flow. Larger windows expose the neural model to longer-term behavior but increase noise, computation, and inference latency. We therefore evaluate windows of 10 and 100 flows.

\paragraph{Inference Query Generation.} 
Each temporal window is converted into a DeepProbLog query containing two complementary inputs. The preprocessed feature vectors form the input to the neural attack-phase classifier, while the current flow's protocol, endpoint location, and service-related attributes are supplied to the symbolic component. The query also identifies the flow under evaluation and requests its most likely attack phase. During inference, DeepProbLog combines the neural phase probabilities with the applicable attack-phase rules and the recorded attack history to produce the final phase distribution.

\section{Neural Ablation Study}
\label{sec:neural_ablation_study}

We conduct an ablation study to select the neural configuration used in the subsequent experiments. For the multiclass and ensemble LSTM models, we vary three factors: the input feature set, temporal window size \(w\), and learning rate \(lr\). The \textit{flow-only} feature set contains the selected Zeek connection attributes, whereas the \textit{base} feature set additionally includes the statistical features computed over recent network activity. The complete set of evaluated configurations is reported in Table~\ref{tab:ablation_experimental_configs}.
\begin{table}[t]
\centering
\fontsize{8}{9.5}\selectfont
\caption{Experimental configurations evaluated in neural baseline models.}
\label{tab:ablation_experimental_configs}
\begin{tabular}{ll}
\toprule
Configuration & Evaluated options \\
\midrule
Feature sets   & Base features, flow-only features \\
Window sizes   & 10, 100 \\
Learning rates & $10^{-3}$, $10^{-4}$ \\
\bottomrule
\end{tabular}
\end{table}

Each configuration is evaluated using five-fold cross-validation with early stopping. To mitigate the substantial class imbalance in both datasets, training batches are constructed using a weighted sampler. We report macro F1-score, accuracy, detection rate (DR), and false alarm rate (FAR), and rank configurations by macro F1-score.

\paragraph{AIT-LDSv2.}
Table~\ref{tab:ablation_ait} reports the eight highest-ranked configurations on AIT-LDSv2. The ensemble LSTM using the base feature set, \(w=100\), and \(lr=10^{-3}\) achieves the best result across all metrics, with a macro F1-score of \(0.8650\). The corresponding multiclass model ranks second. More generally, the strongest configurations use the base feature set and the longer temporal window, indicating that statistical features and extended flow history are beneficial for this dataset.

\begin{table}[t]
\centering
\fontsize{8}{9.5}\selectfont
\caption{Top neural configurations on AIT-LDS V2, ranked by macro F1-score. The best value for each metric is shown in bold.}
\label{tab:ablation_ait}
\begin{tabular}{llcccccc}
\toprule
Model & Features & Window & Learning rate & Macro F1 & Accuracy & DR & FAR \\
\midrule
LSTM\textsubscript{Ensemble}   & Base      & 100 & $10^{-3}$ & \textbf{0.8650} & \textbf{0.9966} & \textbf{0.9874} & \textbf{0.0029} \\
LSTM\textsubscript{Multiclass} & Base      & 100 & $10^{-3}$ & 0.8213 & 0.9952 & 0.9865 & 0.0043 \\
LSTM\textsubscript{Ensemble}   & Base      & 100 & $10^{-4}$ & 0.7498 & 0.9907 & 0.9867 & 0.0090 \\
LSTM\textsubscript{Multiclass} & Base      & 100 & $10^{-4}$ & 0.6919 & 0.9874 & 0.9846 & 0.0124 \\
LSTM\textsubscript{Multiclass} & Base      & 10  & $10^{-3}$ & 0.6749 & 0.9923 & 0.9756 & 0.0067 \\
LSTM\textsubscript{Ensemble}   & Base      & 10  & $10^{-3}$ & 0.6741 & 0.9924 & 0.9760 & 0.0066 \\
LSTM\textsubscript{Multiclass} & Flow-only & 100 & $10^{-3}$ & 0.6367 & 0.9307 & 0.9706 & 0.0715 \\
LSTM\textsubscript{Ensemble}   & Base      & 10  & $10^{-4}$ & 0.6365 & 0.9880 & 0.9742 & 0.0112 \\
\bottomrule
\end{tabular}
\end{table}

\paragraph{DARPA 2000.}
Table~\ref{tab:ablation_darpa} presents the corresponding results for DARPA 2000. The ensemble LSTM using the base feature set, \(w=10\), and \(lr=10^{-3}\) obtains the highest macro F1-score and accuracy. The multiclass model with the same configuration ranks second and achieves the lowest FAR, while the flow-only ensemble obtains the highest DR. In contrast to AIT-LDSv2, the best DARPA 2000 configurations predominantly use the shorter temporal window, suggesting that the relevant attack behavior is captured within a more compact flow history.

\begin{table}[t]
\centering
\fontsize{8}{9.5}\selectfont
\caption{Top neural configurations on DARPA 2000, ranked by macro F1-score. The best value for each metric is shown in bold.}
\label{tab:ablation_darpa}
\begin{tabular}{llcccccc}
\toprule
Model & Features & Window & Learning rate & Macro F1 & Accuracy & DR & FAR \\
\midrule
LSTM\textsubscript{Ensemble}   & Base      & 10  & $10^{-3}$ & \textbf{0.9152} & \textbf{0.99958} & 0.99957 & 0.00030 \\
LSTM\textsubscript{Multiclass} & Base      & 10  & $10^{-3}$ & 0.9142 & 0.99957 & 0.99953 & \textbf{0.00027} \\
LSTM\textsubscript{Ensemble}   & Flow-only & 10  & $10^{-3}$ & 0.9131 & 0.99955 & \textbf{0.99959} & 0.00041 \\
LSTM\textsubscript{Multiclass} & Flow-only & 10  & $10^{-3}$ & 0.8850 & 0.99949 & 0.99955 & 0.00046 \\
LSTM\textsubscript{Ensemble}   & Base      & 100 & $10^{-3}$ & 0.8843 & 0.99955 & 0.99953 & 0.00033 \\
LSTM\textsubscript{Ensemble}   & Flow-only & 100 & $10^{-3}$ & 0.8817 & 0.99953 & 0.99951 & 0.00030 \\
LSTM\textsubscript{Multiclass} & Flow-only & 100 & $10^{-3}$ & 0.8793 & 0.99951 & 0.99953 & 0.00033 \\
LSTM\textsubscript{Ensemble}   & Flow-only & 10  & $10^{-4}$ & 0.8783 & 0.99944 & 0.99953 & 0.00055 \\
\bottomrule
\end{tabular}
\end{table}

\paragraph{Selected Configurations.}
Based on macro F1-score, we select the base feature set and a learning rate of \(10^{-3}\) for both datasets. A window of \(100\) flows is used for AIT-LDSv2, whereas a window of \(10\) flows is used for DARPA 2000. These dataset-specific settings are applied consistently to the neural baselines and to the corresponding neural components of the \name{} models.

\end{document}